\documentclass{vgtc}   

\ifpdf
  \pdfoutput=1\relax                   
  \usepackage{graphicx}                
  \DeclareGraphicsExtensions{.pdf,.png,.jpg,.jpeg} 
\else
  \usepackage{graphicx}                
  \DeclareGraphicsExtensions{.eps}     
\fi%

\graphicspath{{figures/}{pictures/}{images/}{./}} 

\usepackage{microtype}                 
\PassOptionsToPackage{warn}{textcomp}  
\usepackage{textcomp}                  
\usepackage{mathptmx}                  
\usepackage{times}                     
\usepackage{cite}                      
\usepackage{tabu}                      
\usepackage{booktabs}                  

\usepackage{color,soul}

\usepackage{stfloats}

\usepackage{graphicx}
\usepackage{subfigure}
\usepackage{stackengine}
\usepackage{rotating}
\usepackage{tabularx}
\usepackage{amsmath,bm}
\usepackage{float}

\usepackage[toc,page]{appendix}

\onlineid{1024}

\vgtccategory{Research}

\vgtcinsertpkg

\title{A Machine Learning Approach for \\ Predicting Human Preference for Graph Layouts}
\author{Shijun Cai\thanks{e-mail: scai5619@sydney.edu.au} %
\and Seok-Hee Hong\thanks{e-mail: seokhee.hong@sydney.edu.au} %
\and Jialiang Shen\thanks{e-mail: jshe9143@uni.sydney.edu.au} %
\and Tongliang Liu\thanks{e-mail: tongliang.liu@sydney.edu.au}}
\affiliation{\scriptsize University of Sydney, Australia}

\abstract{Understanding what graph layout human prefer and why they prefer is significant and challenging due to the highly complex visual perception and cognition system in human brain. In this paper, we present the first machine learning approach for predicting human preference for graph layouts.

In general, the data sets with human preference labels are limited and insufficient for training deep networks. To address this, we train our deep learning model by employing the transfer learning method, e.g., exploiting the quality metrics, such as shape-based metrics, edge crossing and stress, which are shown to be correlated to human preference on graph layouts. 
Experimental results using the ground truth human preference data sets show that our model can successfully predict human preference for graph layouts.
To our best knowledge, this is the first approach for predicting qualitative evaluation of graph layouts using  human preference experiment data.
} 

\begin{document}

\maketitle

\section{Introduction} 

Evaluation of graph layouts is a significant problem in graph drawing.
A number of quality metrics (or {\em aesthetic criteria}), such as edge crossings, bends, drawing area, total edge lengths, angular resolution, and stress, have been proposed for the {\em quantitative} evaluation of graph layouts~\cite{di1999graph}. 
Consequently, various graph drawing algorithms to optimize these metrics have been developed~\cite{di1999graph}.

{\em Qualitative} evaluation on graph layouts is also available, using HCI methodology, using {\em human preference} or  specific {\em task performance}, measuring time and error. 
For example, edge crossings are shown to be important aesthetic criteria for performing human preference and shortest path tasks on graph layouts~\cite{purchase97aesthetic,ware2002cognitive}.
Furthermore, large crossing angles are shown to be effective for shortest path tasks on graph layouts, when edge crossings are present~\cite{huang2008effects}.

Understanding what layout human prefer and why they prefer a specific graph layout over the others is significant, since it motivates researchers to design algorithms to compute such layouts, and guides users to choose specific algorithms to produce such layouts.
However, it is extremely challenging due to the highly complex human visual perception and cognition system involving massively parallel processing using vision and memory in human  brain~\cite{ware2019information}.

A series of {\em human preference} task performance experiments have been conducted to better understand which graph layout human prefer.
For example,  Purchase~\cite{purchase97aesthetic} found the correlation between the human preference and fewer edge crossings in graph layouts.
More recently, Chimani et al.~\cite{chimani2014people} found the correlation between the human preference and lower stress in graph layouts, and Eades et al.~\cite{eades2015shape} found the correlation between the human preference and higher shape-based metrics in graph layouts.

In this paper, we present the first deep learning approach for predicting human preference for graph layouts. 
More specifically, we propose a CNN-Siamese-based model that can be trained to predict human preference  from a given pair of layouts of the same graph.  
Roughly speaking, CNN models read images rendered from graph layouts and 
convert them into feature vectors, which are inspired by the processing procedure of the human brain~\cite{goodfellow2016deep}.
The Siamese model computes the difference between a pair of feature vectors.

To understand the procedure, we assume that there exists a certain human preference measurement which measures a pair of layouts to a human preference label (that indicates which layout human prefer). The deep model can be regarded as a measurement from the input layout pair to the output prediction. 
The model is trained to mimic the ground truth human preference measurement by fitting the training data which contains pairs of layouts and human preference labels.

The amount of the ground truth human preference data sets from previous experiments~\cite{chimani2014people,eades2015shape} is relatively limited and insufficient for training deep models. 
To address this,
we train the model by employing the {\em transfer learning} method \cite{pan2009survey}, which exploits data from related problems to help the original problem (e.g., human preference prediction). 
Specifically, since some quality metrics (i.e., shape-based metrics, edge crossing and stress metrics) have been shown to be correlated to human preference for graph layouts~\cite{chimani2014people,eades2015shape}, we {\em transfer the hypothesis} by employing the layout pairs labeled by quality metrics to help train our deep model. 

More specifically,
we first pre-train our model using {\em quality-metrics-based pairs} described in Section~\ref{sec:metric}, which consists of different graph layouts of the same graph and labels constructed by employing the quality metrics. 
Then, we fine-tune our model by using the human preference experiment data. 
Extensive experiments show that our machine learning approach successfully predicts human preference for graph layouts.

The main contribution of this paper is summarized as follows:
\begin{enumerate}
\setlength{\itemsep}{0pt}
\setlength{\parsep}{0pt}
\setlength{\parskip}{0pt}
\item 
We propose the first machine learning approach to predict human preference for graph layouts. To our best knowledge, this is the first approach for predicting qualitative evaluation of graph layouts by exploiting the ground truth human preference experiment data~\cite{chimani2014people,eades2015shape}.

\item 
We introduce a transfer learning method to overcome the insufficiency 
of the available human preference experiment data for training deep models.
More specifically, we use quality-metrics-based pairs to pre-train our model to better understand human preference.

\item 
Extensive experiments using the ground truth human preference data~\cite{chimani2014people,eades2015shape} show that our machine learning approach successfully predicts human preference for graph layouts.
Specifically, our model achieves average test accuracy of $92.28\%$ for large graphs, and $63.77\%$ for small graphs, significantly outperforming random guessing (i.e., greater than $50\%$) for the {\em binary} human preference problem. 

\end{enumerate}

The main reason of the high test accuracy for large graphs (e.g., mesh and large scale-free graphs) is due to the strong preference scores among the layouts in the human preference data (i.e., some layouts have much better quality than other layouts).  

On the other hand, for small graphs (e.g., small sparse graphs and biconnected graphs), there was no strong preference among the layouts in the human preference data (i.e., all layouts have similar good quality).
Therefore, it is more difficult to predict. 


\section{Related Work}

\subsection{Quantitative Evaluation for  Graph Drawing}

Various {\em quality metrics} for the evaluation of graph drawings, called {\em aesthetic criteria}, are available~\cite{di1999graph}.
Traditional {\em readability} metrics include edge crossings, bends, area, total edge lengths and angular resolution.
Consequently, many graph drawing algorithms have been designed to optimize these quality metrics~\cite{di1999graph}.

Recently, new {\em faithful}  metrics have been developed, which measure how faithfully graph drawings visually display the ground truth structures of graphs.  
For example, Eades et al.~\cite{eades2015shape} introduced
the {\em shape-based metrics}, by comparing the similarity between a graph $G$ with a proximity graph $G'$ computed from a drawing of $G$. 
The {\em stress}~\cite{di1999graph} is a  {\em distance faithful} metrics, which compare the difference between graph theoretic distance of vertices and the Euclidean distance in a drawing.

Similarly, the \textit{cluster faithful} metrics~\cite{amyragd} compare the similarity between the ground truth clustering of a graph $G$ and the geometric clustering computed from a drawing of $G$.
The \textit{symmetry faithful} metrics~\cite{amyrapvis} measure how the ground truth {\em automorphisms} of a graph are displayed as symmetries in a drawing.

\subsection{Qualitative Evaluation for  Graph Drawing}

Qualitative evaluation on graph layouts have been investigated by conducting the HCI-style human experiments, with preference or task performance, measuring time and error.
For example, Purchase~\cite{purchase97aesthetic} found correlation between human preference and fewer edge crossings in graph layouts.

Ware~\cite{ware2002cognitive} found correlation between fewer edge crossings and shortest path tasks in graph layouts, and 
Huang et al.~\cite{huang2008effects} found correlation between large {\em crossing angles} and the shortest path task. 

Recent studies~\cite{marner2014gion} found that human {\em untangling} interaction task of hairball-like graph layouts is positively correlated with the shape-based metrics~\cite{eades2015shape}, while surprisingly negatively correlated with the edge crossings and stress.

\subsection{Human Preference Experiments in~\cite{chimani2014people,eades2015shape}}

More recently, a series of  human preference experiments have been conducted~\cite{chimani2014people,eades2015shape}. 
Specifically, in the human preference experiments, the system showed two layouts of the same graph, randomly chosen from five different graph layouts, including  force-directed layouts (such as FR~\cite{fruchterman1991graph}), stress minimization layouts and multi-level layouts (such as FM3~\cite{hachul2004drawing}). 
The data set used in the experiment includes Hachul library, Walshaw’s Graph Partitioning Archive, and randomly generated biconnected and scale-free graphs.
The task for participants was to choose their preferred layout from a pair of layouts of the same graph,
and select their {\em preference score} using a slider bar scaled from 0 to 5.

The first experiment 
conducted at the University of Osnabrück~\cite{chimani2014people}
found the correlation between human preference for graph layouts and  {\em edge crossings} and {\em stress}. Namely, human prefer graph layouts with less stress and fewer crossings. 

The two follow-up experiments~\cite{eades2015shape} 
conducted at the Graph Drawing conference 2014 and the University of Sydney, 
showed that the {\em shape-based metrics} are positively correlated with human preference, i.e., human prefer graph layouts with high shape-based 
metrics.

\subsection{Deep Learning}

Recently, deep learning has achieved great success in various fields, such as  computer vision, natural language processing,
and speech recognition. 
The {\em Convolutional Neural Network}  (CNN) is a representative deep network for image recognition and classification. 
CNNs are a type of multi-layer neural networks, designed and trained to recognize the nature of images by varying the depth and breadth of a model~\cite{krizhevsky2012imagenet}. 

After the AlexNet introduced by Krizhevsky et al.~\cite{krizhevsky2012imagenet}, much deeper and more complex CNNs has been developed, such as  VGG~\cite{simonyan2014very}, GoogLeNet/Inception~\cite{ioffe2015batch} and ResNet-50~\cite{he2016deep}. 

{\em Siamese neural networks} were  introduced by Bromley and LeCun to solve signature verification as an image discrimination problem~\cite{bromley1994signature}. 
Koch et al.~\cite{koch2015siamese} used Siamese neural networks to rank the similarity between multiple inputs and discriminated input features. 
A Siamese neural network joins together the highest-level feature representations of twin inputs for image classification problems.

{\em Transfer learning} \cite{pan2009survey} aims to improve the learning performance of a target task (or problem) by borrowing knowledge from related but different tasks. Its core idea is to learn task-invariant data representations \cite{gong2016domain}. In computer vision, complex deep networks, e.g., AlexNet~\cite{krizhevsky2012imagenet} and  VGG~\cite{simonyan2014very}, are often trained by employing the transfer learning technique to leverage the large-scale dataset \textit{ImageNet} \cite{deng2009imagenet}. 
Specifically, the networks are usually pre-trained on ImageNet first and then are fine-tuned on the datasets of the target tasks.

\subsection{Deep Learning Approaches in Graph Visualization}

A number of researchers used deep learning methods for problems in graph visualization, mainly focusing on quality metrics~\cite{klammler2018aesthetic, haleem2019evaluating, kwon2019deep, wang2019deepdrawing}.
For example, Haleem et al.~\cite{haleem2019evaluating} used CNNs to predict multiple readability metrics, e.g., node spread, group overlap and edge crossings, using graph layout images with up to 600 vertices. 

Klammler et al. \cite{klammler2018aesthetic} used the Siamese neural network for aesthetic discrimination of paired layouts, where the labeled training pairs consisted of a graph layout $D$ and its {\em deformed layout} $D'$ using deformation steps
(i.e., comparing the original layout with its deformed layout using {\em quality metrics}). 
Therefore, their work predicted a better quality  layout based on {\em quantitative evaluation} (i.e., quality metrics).

Note that in our work, we use the ground truth human preference experiment data~\cite{chimani2014people,eades2015shape} for training and testing, where a pair of layouts $D_1$ and $D_2$ are computed using two different graph layout algorithms (i.e., comparing {\em two different graph layouts} for {\em human preference}).
Therefore, We predict human preference based on {\em qualitative evaluation} (i.e., human preference experiment data).

\section{A CNN-Siamese-based Approach}

\subsection{Model}

\begin{figure*}[!ht]
\includegraphics[width=\textwidth]{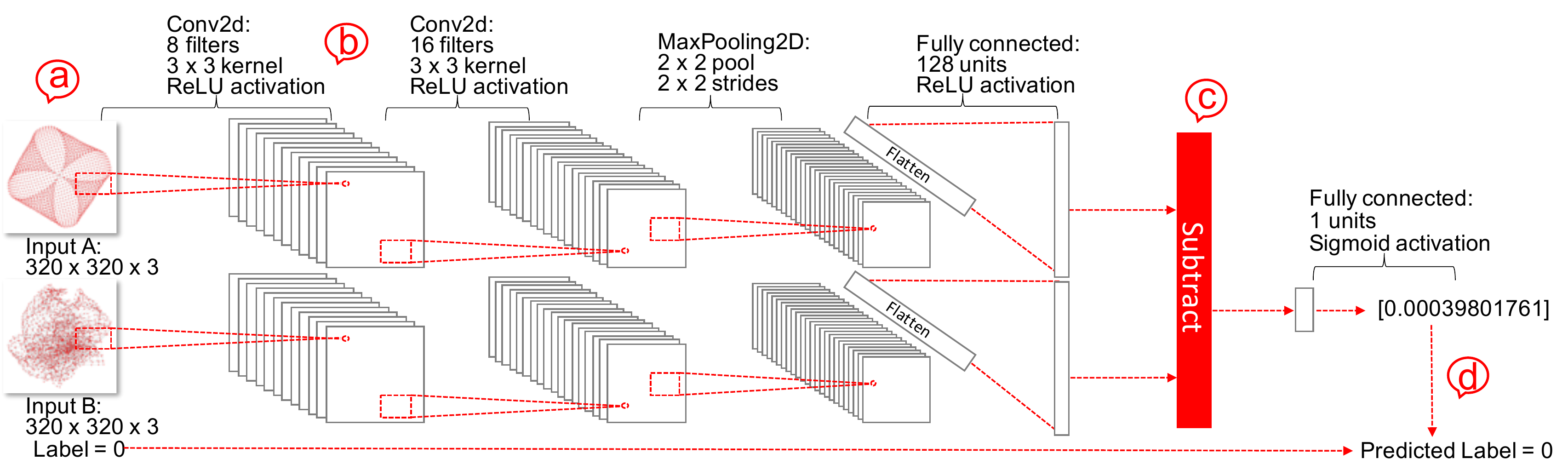}
\vspace{-9mm}
\caption{Our CNN-Siamese-based model: (a) Input data, (b) Twin CNN-based image feature extractor, (c) Subtract part of the Siamese model and (d) Output prediction.} \label{model}
\vspace{-5mm}
\end{figure*}

We present a CNN-Siamese-based model that can predict which layout human prefer from a given pair of layouts. 
The notable advantage of CNNs is that they are powerful in extracting features from image inputs. 
The use of the Siamese model is natural since it deals with a pair of layouts to measure their difference. 

\autoref{model} shows the pipeline of our model, including four essential parts: 
(a) Input data, i.e., a pair of layouts (note that in the training phase, the input pair also needs a {\em label} indicating which layout is better; while in the test phase, the label is not required); 
(b) Twin CNN-based image feature extractors; 
(c) Subtraction part of the Siamese model; 
(d) Output prediction (note that in the training phase, we need to measure and minimize the difference between the predicted value and the ground truth; while in the test phase, the output indicates the predicted human preference label). 
We now explain each part of the model in detail.

\subsubsection*{(a) Input data}

To obtain a machine learning algorithm for predicting human preference, we need training and validation data sets to train the parameters.
Furthermore, we need to select a model  
(controlled by hyper-parameters that cannot learn from the training data) with good performance. 
Specifically, training and validation data sets consist of labeled pairs of color images.
Such images are the visual rendering of layouts with size $320\times320$.
We label each pair with 0 or 1 during training,
where 0 means 
human prefer the first layout A compared with the second layout B (see~\autoref{model}(1)); otherwise, human prefer the second layout B.
More details about the training and validation data sets are presented in~Section~\ref{sec:experiments}. 

Note that the data with human preference labels is limited. 
We employ transfer learning and the data pairs (image pairs) labeled by quality metrics to help train our model. 
More information about transfer learning and the data is presented in~Section~\ref{sec:transfer_learning}.

\subsubsection*{(b) Twin CNN-based image feature extractor}
Our twin CNN-based image feature extractor is built on VGG. 
They convert the input images into semantic feature vectors.
The following part of the Siamese model outputs a prediction for human preference based on the semantic feature vectors. 
We now explain the details of our CNN-based feature extractor.


{\em Convolutional layers} are efficient to extract semantic features of an image inspired by the processing procedure of human brain.
Multiple {\em hidden layers} are essential to increase the expressive ability of the deep network.
{\em Max pooling layers} retain the most significant features.
The {\em  Fully connected layer} summarizes features for feature subtraction in the Siamese model (see~\autoref{model}(c)). \autoref{model}(b) shows the design of our CNN-based feature extractor. 

\subsubsection*{(c) Subtraction part of the Siamese model}
A Siamese neural network consists of twin feature extractors and a joint layer. To compute the difference of twin inputs, we use a subtraction part to join the twin features, which shows a higher test accuracy than the concatenation part.
The subtraction part converts the pair of semantic feature vectors output by the twin feature extractors into a single value in the range $[0,1]$ to predict human preference. 
Specifically, the two feature vectors are joined into a single long vector, which is reduced to one dimensional by employing a fully connected layer. 
A Sigmoid activation function further restricts the output value in the range $[0,1]$.
Note that if the output value $\leq 0.5$,
the prediction is that 
human prefer the first layout A compared with the second layout B; vice versa.

\subsubsection*{(d) Output prediction} 
We aim to train the proposed deep model to have outputs that are aligned with human preference labels (see~\autoref{model}(d)). 
Specifically, to achieve this, we optimize the parameters of the deep model by minimizing the difference between the predicted values and the ground truth human preference labels. 
The difference can be measured by a loss function. 
In this paper, we employ the binary cross-entropy loss function.

Since the training sample with human preference could be limited, we employ the transfer learning technique to reduce the hypothesis complexity of the deep model.



\subsection{Transfer Learning for Predicting Human Preference for Graph Layouts}\label{sec:transfer_learning}

To train a deep machine to understand human preference, we need a large amount of human-labeled pairs of layouts. 
However, annotating a large number of layout pairs is usually time-consuming and expensive. 
Fortunately, we can address the issue by employing the transfer learning technique~\cite{pan2009survey}.
Transfer learning transfers knowledge across different tasks to improve learning performance. 
Typically, if the target task has limited training data, by using transfer learning, we could use the related tasks (called source tasks) that have sufficient training data.

It has been shown that some quality metrics (i.e., shape-based metrics, edge crossing and stress metrics) are correlated to human preference~\cite{chimani2014people,eades2015shape}.
The layout pairs with different quality metrics values are relatively easy to compute, which means we could easily obtain example pairs labeled by quality metrics.
We thus exploit such data to help training our deep model. 
More details on how to generate those data are described in Section~\ref{sec:labeled pairs}.

Because of the highly complex visual perception and cognition system in the human brain, the mechanism that human use to decide their preference for graph layouts could be very complicated. 
Although some quality metrics are shown to be correlated to human preference, the precise relationship between them remains unknown. 
To address the above issue, we {\em transfer the hypothesis}. Specifically, using the related data (e.g., quality-metrics-based pairs) to pre-train the deep model and then use the target task data (e.g., human preference pairs) to fine-tune the deep model.


To demonstrate the effectiveness of the transfer learning, we compare the following three trained models in our experiments as an ablation study:
\begin{enumerate}
\setlength{\itemsep}{0pt}
\setlength{\parsep}{0pt}
\setlength{\parskip}{0pt}
    \item {\em M}: a model trained only on data pairs labeled by using quality metrics. 
    \item {\em HP}: a model trained only on data pairs with human preference labels.
    \item {\em M+HP}: a model pre-trained on data pairs labeled by using quality metrics and then fine-tuned on data pairs with human preference labels (i.e., a model trained by using the first and second stages).
\end{enumerate}
Note that M and HP are two baseline models to compare with our transfer learning model, i.e., M+HP. 
In the experiment section, results show that the performance of M+HP is much better than these of HP and M.

\subsection{Generating Labeled Pairs}\label{sec:labeled pairs}


\subsubsection{Human preference pairs}\label{sec:HP}
Human preference pairs are processed from data in the human preference experiments~\cite{chimani2014people,eades2015shape}.
In the previous experiments, given a pair of layouts of the same graph, 
the task for multiple participants are to choose their preferred layout. The task performance is measured by a discrete preference score ranging from $0$ to $5$, 
where $0$ means no preference for the pair of layouts, and $5$ means the strongest preference score. 
Note that since human preference may be subjective, different participants may have different preferences for the same pair of layouts. 
To solve this conflict, we use the weighted majority voting methods to reach a consensus.

Specifically, let $G_{i}$ denote the graph where $i$ is graph index. Let $D_{i\_j}$ and $D_{i\_k}$ denote two layouts of the graph $G_i$ where $j$ and $k$ are the layouts indices. 
Let $P$ denote the preferred layout. 
Let $S$ be the preference score of the preferred layout. 
For the pair $D_{i\_j}$ and $D_{i\_k}$, where $j<k$, 
we convert the measurement of the task performance into the {\em human preference label (e.g., 0 or 1)}, reflecting the consensus of multiple preferences of the same pair of layouts.
The human preference label is generated by the following weighted majority voting algorithm:
\begin{enumerate}
\setlength{\itemsep}{0pt}
\setlength{\parsep}{0pt}
\setlength{\parskip}{0pt}
    \item Counting how many participants chose their preference for the pair $D_{i\_j}$ and $D_{i\_k}$. Let $N$ denotes the number of participants.
    \item Assigning the weights for the participants to vote. For the $n$-th participant, if the preferred layout $P$ is $D_{i\_j}$, set its voting weight $w_n$ to be the preference score of the preferred layout, i.e., $S$; if $P$ is $D_{i\_k}$, set its voting weight $w_n$ to be $-S$.
    \item Labeling the pair according to the weighted majority voting. If the average preference score $\sum_n{w_n}/n>0$, set the label $L_{hp}$ to be 0, 
    indicating that the layout $D_{i\_j}$ is more preferred by human than the layout $D_{i\_k}$;
    if $\sum_n{w_n}/n<0$, set the label $L_{hp}$ to be 1, 
    indicating that the layout $D_{i\_j}$ is less preferred than the layout $D_{i\_k}$; 
    for the pair $D_{i\_j}$ and $D_{i\_k}$, if $\sum_n{w_n}/n=0$, discard the pair of layouts without labeling.
\end{enumerate}

\subsubsection{Quality-metrics-based pairs}\label{sec:metric}

To train and evaluate our model, 
we randomly split graphs and having their corresponding labeled layout pairs into two data sets, i.e., a {\em training data set} and a {\em test data set}. 
However, the size of layout pairs with human preference labels can be small, making it hard to train deep models well. 
We handle this problem by employing the transfer learning technique to pre-train the deep models, which exploits the layout pairs labeled by using quality metrics. 

For a pair of layouts, it is labeled by exploiting shape-based metrics $M_{sh}$, edge crossing $M_c$, and stress metrics $M_{st}$. 
The works of literature~\cite{chimani2014people, eades2015shape} show that values of shape-based metrics are positively correlated to human preference; while values of edge crossing and stress metrics are negatively correlated to human preference. 
Note that quality metrics for layouts can be computed by specific algorithms~\cite{purchase2002metrics, eades2015shape}.
As there are different quality metrics, we form quality-metrics-based labels by employing the majority voting method.


In the human preference experiments~\cite{eades2015shape}, each graph has five layouts.
We compute the quality-metrics-based labels for all possible ten pairs of each graph.
Specifically, let $D_{i\_j}$ and $D_{i\_k}$ be two layouts of the graph $G_i$, the quality-metrics-based label is generated as follows:
\begin{enumerate}
\setlength{\itemsep}{0pt}
\setlength{\parsep}{0pt}
\setlength{\parskip}{0pt}
    \item Computing the quality metrics values for the layouts $D_{i\_j}$ and $D_{i\_k}$. Let use $M_{sh\_j}$, $M_{c\_j}$ and $M_{st\_j}$ (resp. $M_{sh\_k}$, $M_{c\_k}$ and $M_{st\_k}$) to denote the values of the shape-based metric, edge crossing, and stress values of layout $D_{i\_j}$ (resp. $D_{i\_k}$).
    \item Assigning intermediate labels. 
    Let $L_{sh}$, $L_{c}$, and $L_{st}$ be the intermediate labels induced by the three quality metrics. 
    Set $L_{sh} =0$ (resp. $L_{c} =0$ and $L_{st} =0$), if $M_{sh\_j} > M_{sh\_k}$ (resp. $M_{c\_j} < M_{c\_k}$ and $M_{st\_j} < M_{st\_k}$).
    Set $L_{sh} =1$ (resp. $L_{c} =1$ and $L_{st} =1$), if $M_{sh\_j} < M_{sh\_k}$ (resp. $M_{c\_j} > M_{c\_k}$ and $M_{st\_j} > M_{st\_k}$).
    Discard the label, if $M_{sh\_j} = M_{sh\_k}$ (resp. $M_{c\_j} = M_{c\_k}$ and $M_{st\_j} = M_{st\_k}$).
    \item Labeling the pair according to the majority voting method by using the intermediate labels, 
    e.g., if the majority intermediate labels are 0, set the final quality-metrics-based label for the pair as 0, indicating that the layout $D_{i\_j}$ is more preferred than the layout $D_{i\_k}$ by just exploiting the quality metrics.
\end{enumerate}




\section{Experiments}\label{sec:experiments}

\subsection{Data Set}
In our experiment, 146 graphs and their five layouts from the human preference experiments~\cite{eades2015shape} were used. 
The graphs range in size from small (25 vertices and 29 edges) to large (8,000 vertices and 118,404 edges), 
and have different structures such as sparse graph (i.e., graphs consists of cycles and tree), biconnected graph, mesh, and large scale-free graph. 
We categorized these graphs based on the number of vertex, i.e., large graphs (e.g., mesh and large scale-free graphs in~\autoref{pairs_mh}), and small graphs (e.g., sparse graphs and biconnected graphs in~\autoref{pairs_cb}).
\autoref{graphs} shows the statistics of the graph data used in our experiment.

\begin{table}[h]
\centering
\small
\begin{tabular}{|c|m{0.07\textwidth}<{\centering}|c|c|c|}
\hline
Category&Type&$|V|$&$|E|$&$D$\\\hline
\emph{small}&\emph{sparse}&25 - 363&29 - 468&1.00 - 1.50\\\hline
\emph{small}&\emph{biconnected}&34 - 240&78 - 477&1.92 - 2.94\\\hline
\emph{large}&\emph{mesh}&397 - 8,000&729 - 15,580&1.41 - 1.95\\\hline
\emph{large}&\emph{large scale-free}&1,647 - 5,452&4,769 - 118,404&2.30 - 21.72\\\hline
\end{tabular}
\vspace{1mm}
 \caption{The statistics of the data used in our experiments. The column $|V|$ (resp. $|E|$ and $D$) shows the range of the number of vertex (resp. edge and density) for each type of graph.  } 
\label{graphs}
\vspace{-5mm}
\end{table}

After we pre-process human preference data, as described in~Section~\ref{sec:HP}, we obtain 511 layout pairs with human preference labels.
We then compute the ten quality-metrics-based pairs of each graph using the approach described in~Section~\ref{sec:metric}. 
Therefore, in total, we have 1,460 quality-metrics-based pairs for pre-training, and 511 human preference pairs for fine-tuning and testing.

\subsection{Model Training}
We randomly split the data with human preference labels into training and test sets with a ratio of $7:3$. 
Note that the quality-metrics-based pairs are only used for training, i.e., pre-training. 
Specifically, to split the data, we randomly split the graphs and use their corresponding labeled layout pairs as the training and test sets. 
In other words, layout pairs belonging to graphs of the training set will not appear in the test set. 

To choose the model,
We use the five-fold cross-validation method. 
Specifically, the training data is randomly split into five folds. 
The model is chosen by alternatively using four folds of data for training and the remaining one fold for testing.


\subsection{Prediction Results}

To examine the variations of human preference on different kinds of graphs, we do the ablation study by dividing our data sets into four subsets according to graph types: 
mesh (see $G_{6}$ and $G_{10}$) and large scale-free graph (see $G_{3}$ and $G_{15}$) in~\autoref{pairs_mh}, 
sparse graph (see $G_{0}$ and $G_{188}$) and biconnected graph (see $G_{18}$ and $G_{65}$) in~\autoref{pairs_cb}. 





\autoref{acc} summarizes our experiment results with three trained models (i.e., M, HP and M+HP) on different graph types. 
The number in each cell represents the average test accuracy with the standard deviation after five times random splitting. 

\begin{table}[h!]
\centering
\small
\begin{tabular}{|m{0.1\textwidth}<{\centering}|c|c|c|}
\hline
Type&\textit{M}&\textit{HP}&\textit{M+HP}\\\hline
\emph{sparse}&$(56.57\;\pm\; 3.12) \%$&$(58.37\;\pm\; 1.32) \%$&$\bm{(62.14\;\pm\; 2.59) \%}$\\\hline
\emph{biconnected}&$(52.49\;\pm\; 2.30) \%$&$(61.30\;\pm\; 3.12) \%$&$\bm{(65.40\;\pm\; 3.71) \%}$\\\hline
\emph{mesh}&$(76.49\;\pm\; 2.76) \%$&$(82.51\;\pm\; 2.92) \%$&$\bm{(86.55\;\pm\; 3.20) \%}$\\\hline
\emph{large scale-free}&$(82.85\;\pm\; 4.73) \%$&$(82.81\;\pm\; 3.67) \%$&$\bm{(98.00\;\pm\; 4.47) \%}$\\\hline
\end{tabular}
\vspace{1mm}
\caption{Test accuracy and standard deviation of three trained models: 
the best test accuracy is $98\%$ with standard deviation $\pm4.47\%$ by model M+HP in large scale-free graphs, followed by $86.55\%$ with standard deviation $\pm3.20\%$ in mesh, showing that our model is effective to predict human preference for graph layouts.}
\label{acc}
\vspace{-7mm}
\end{table}

To validate whether one trained model outperforms another, we employ the Wilcoxon signed-rank test to compare the proposed models, a non-parametric statistical hypothesis test method. 


We run the {\em significance test} on the test accuracy of models trained on different types of graphs (i.e., by scipy.stats.wilcoxon function). 
Table~\ref{sig} shows the $p$-values for comparing training variations.
The $p$-value depends on the median accuracy of the first model that is negative against the median accuracy of the second model that is positive; the smaller the $p$-value, the better the second model.
The $p$-value $\leq 0.05$ means the difference is statistically significant. 

\begin{table}[h!]
    \centering
    \begin{tabular}{|l|c|}
    \hline
    Comparison&$p$-value\\\hline
    \emph{M+HP vs M}&0.00000006\\\hline
    \emph{M+HP vs HP}&0.00004286\\\hline
    \emph{HP vs M}&0.00064817\\\hline
    \end{tabular}
    \vspace{1mm}
 \caption{The $p$-values of the Wilcoxon signed-rank tests for comparing our three proposed models trained on different types of graphs.}
 \label{sig}
\vspace{-6mm} 
\end{table}

\section{Summary, Discussion and  Future Work}

{\bf Summary:} In summary, we show human preference for graph layouts can be predicted by machine learning algorithms. 
Specifically, the proposed model predicts human preference for a pair of graph layouts with an average test accuracy of $92.28\%$ for large graphs, and $63.77\%$ for small graphs, which are significantly better than random guessing for binary human preference problem.
Our main hypothesis that human preference for graph layouts can be learned by machine has been successfully proven.

For the three trained models, i.e., M, HP and M+HP, the performance gradually increases significantly, as shown in~\autoref{acc}, which shows the importance of employing the transfer learning technique during training. 
For details, see Appendix~\ref{sec:Comparison_on_trained_models}.

{\bf Large graphs:}
Large graphs have high test accuracy due to the relatively clear preference between layouts (i.e., some layouts have much better quality than others).  
For mesh, their layouts have distinct shapes, therefore it is easier for human to decide their preference with a high preference score
(see [$D_{5\_4}$, $D_{5\_3}$] in~\autoref{pairs:g1}).
Large scale-free graphs are difficult to visualize, i.e., some graph layouts produce much better quality than other layouts with poor quality, therefore human can choose their preference with a high preference score 
(see [$D_{3\_0}$, $D_{3\_3}$] in~\autoref{pairs:g1}).
For details, see Appendix~\ref{sec:Large_graphs}.

{\bf Small graphs:}
Our three trained models succeed to predict the ground truth human preference value for the small graphs data set (see~\autoref{pairs:g2}).
However, for small graphs, the human preference was not strong between layouts (i.e., layouts have similar quality), 
see~\autoref{pairs:e2} and~\autoref{e2}.
For details, see Appendix~\ref{sec:Small_graphs}.

{\bf Limitation and Future work:}
Although our trained models perform quite well on both large graphs and small graphs, we found some limitations. 
For example, when a pair of layouts has similar visual quality, the preference score is very low, i.e., 0 or 1
(see [$D_{113\_1}$, $D_{113\_2}$] in ~\autoref{pairs:f} and~\autoref{f}).  
Moreover, preference scores of the same pair of layouts can vary  (from 1 to 4) due to the subjective human preference
(see [$D_{13\_4}$, $D_{13\_0}$] in \autoref{pairs:f} and~\autoref{f}).
Interestingly, there was an exceptional case where only model M succeeds to predict human preference
(see~\autoref{pairs:f2} and~\autoref{f2}).

To mitigate such limitations, we plan to conduct new human preference experiments with more data sets and  visually different layouts, to design a machine learning model for better predicting the human preference on  graph layouts.

\acknowledgments{
This work was supported by ARC Linkage Project with Oracle Research lab.}
\vspace{-2mm}


\clearpage
\begin{appendices}
\section{Examples of graphs}\label{sec:pairs}
\begin{figure*}[hb!] 
\newcommand{\tabincell}[2]{\begin{tabular}{@{}#1@{}}#2\end{tabular}}
  \centering
    \begin{tabular}{l c c c c c}
      & $D_{6\_0}$ & $D_{6\_1}$ & $D_{6\_2}$ & $D_{6\_3}$ & $D_{6\_4}$ \\\
        \tabincell{l}{$G_{6}$\\mesh\\$|V|$=397\\$|E|$=760\\$D$=1.91}
        &
          \begin{minipage}{0.15\textwidth}
            \includegraphics[width=\textwidth]{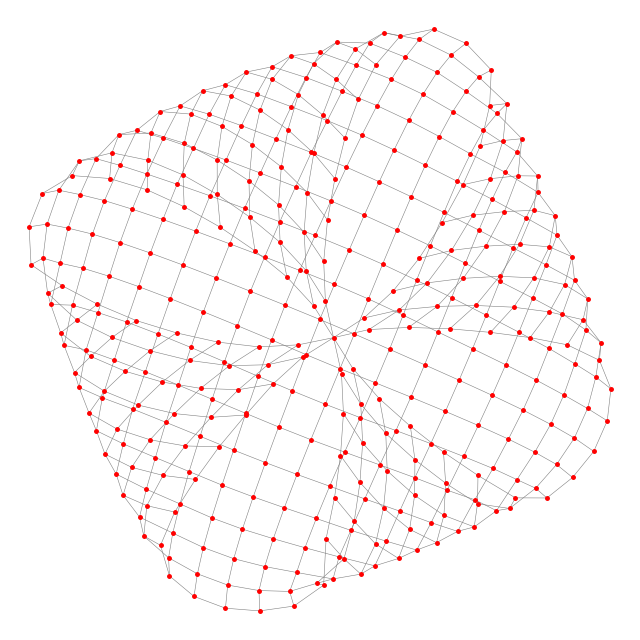}
        \end{minipage}
        &
        \begin{minipage}{0.15\textwidth}
            \includegraphics[width=\textwidth]{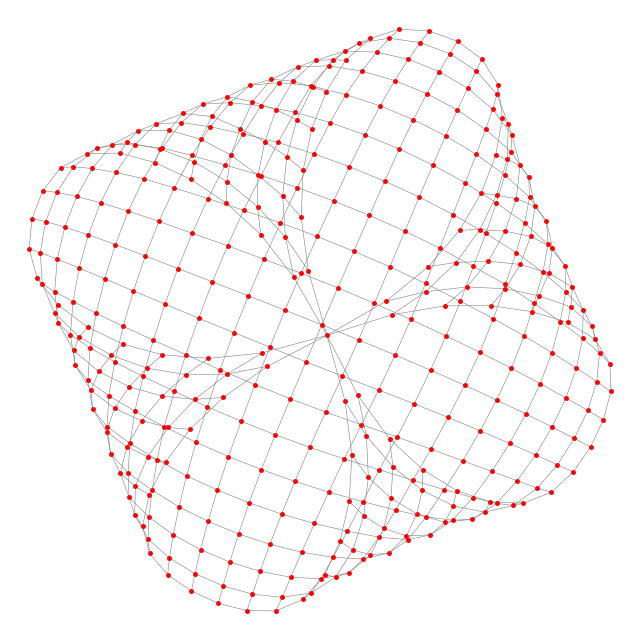}
        \end{minipage}
        &
        \begin{minipage}{0.15\textwidth}
            \includegraphics[width=\textwidth]{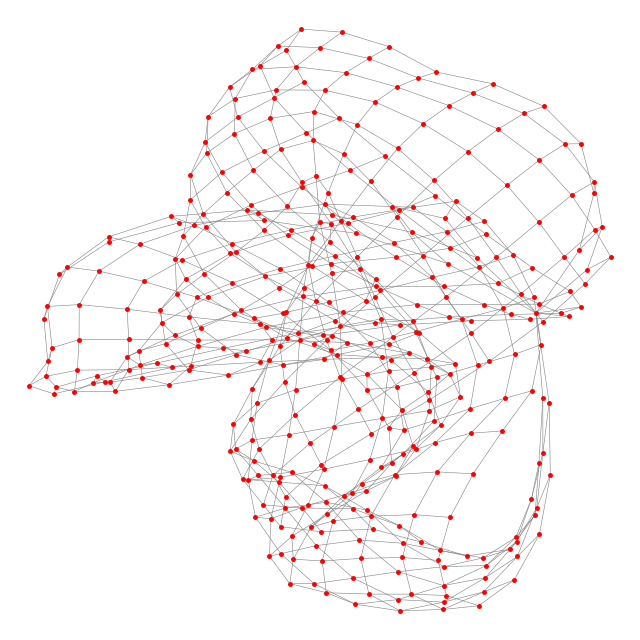}
        \end{minipage}
        &
        \begin{minipage}{0.15\textwidth}
            \includegraphics[width=\textwidth]{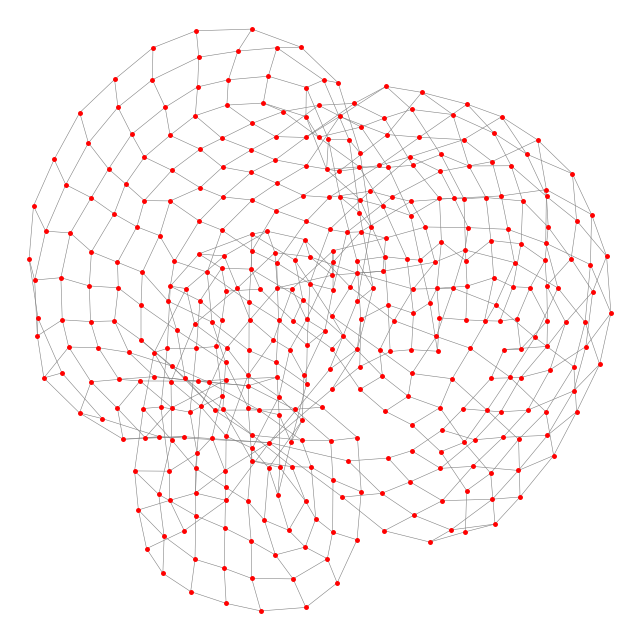}
        \end{minipage}
        &
        \begin{minipage}{0.15\textwidth}
            \includegraphics[width=\textwidth]{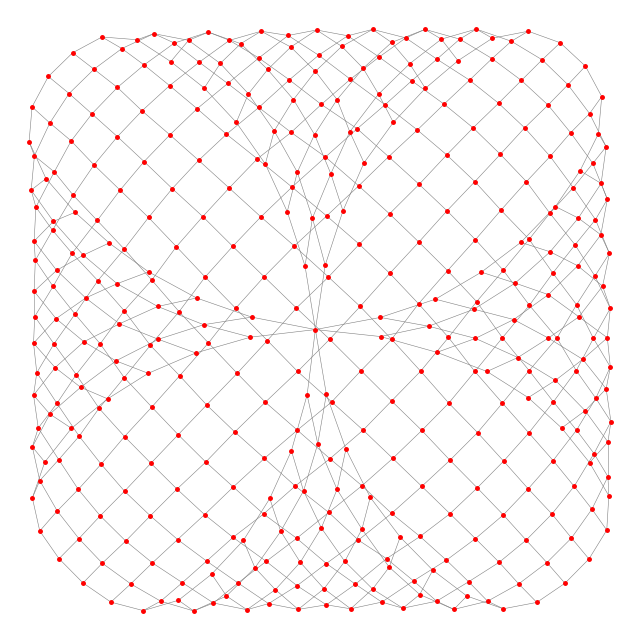}
        \end{minipage}
        \vspace{1mm}
        \\\
      & $D_{10\_0}$ & $D_{10\_1}$ & $D_{10\_2}$ & $D_{10\_3}$ & $D_{10\_4}$
      \\\
        \tabincell{l}{$G_{10}$\\mesh\\$|V|$=1,599\\$|E|$=3,120\\$D$=1.95}
        &
          \begin{minipage}{0.15\textwidth}
            \includegraphics[width=\textwidth]{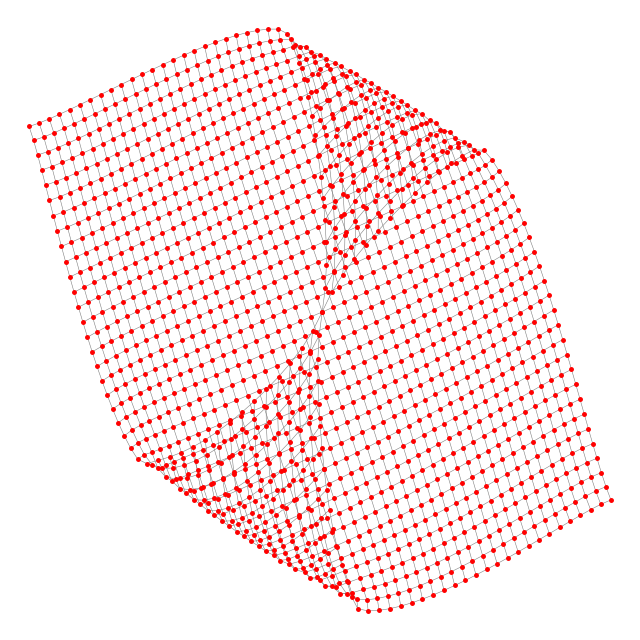}
        \end{minipage}
        &
        \begin{minipage}{0.15\textwidth}
            \includegraphics[width=\textwidth]{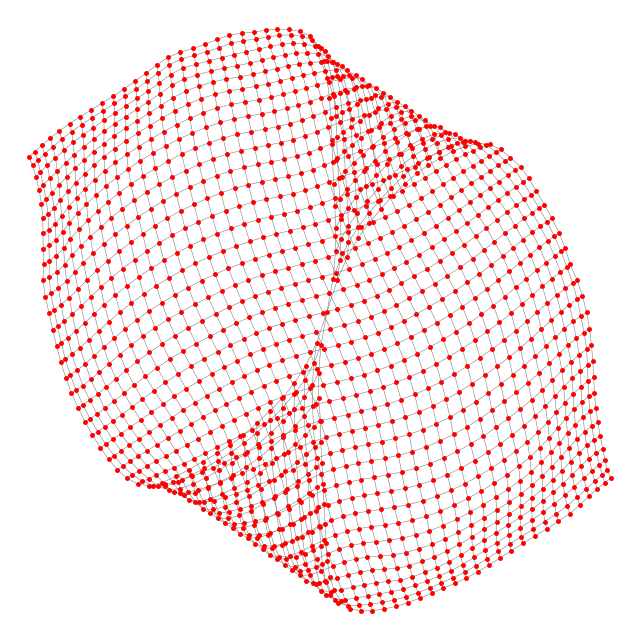}
        \end{minipage}
        &
        \begin{minipage}{0.15\textwidth}
            \includegraphics[width=\textwidth]{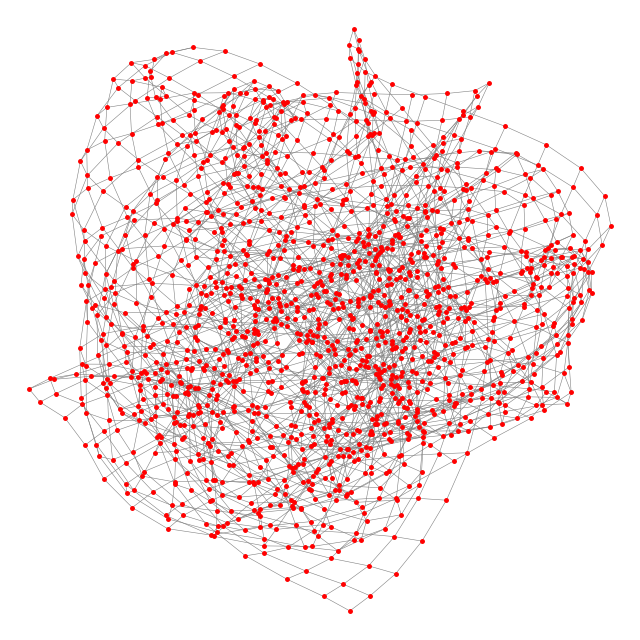}
        \end{minipage}
        &
        \begin{minipage}{0.15\textwidth}
            \includegraphics[width=\textwidth]{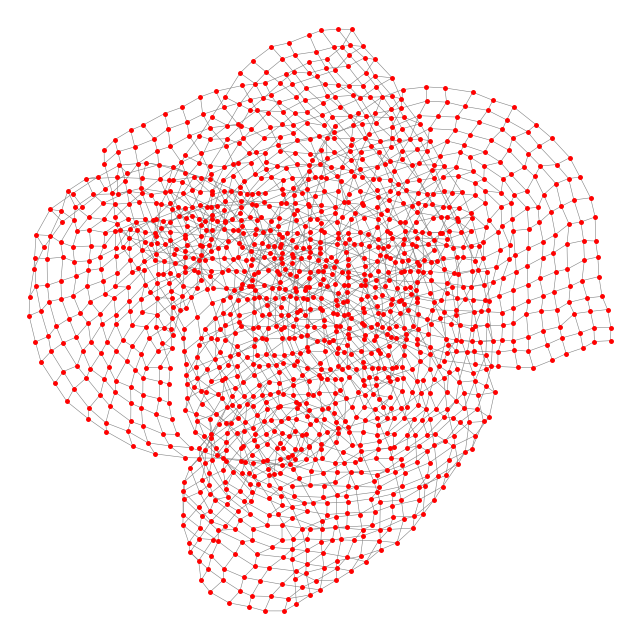}
        \end{minipage}
        &
        \begin{minipage}{0.15\textwidth}
            \includegraphics[width=\textwidth]{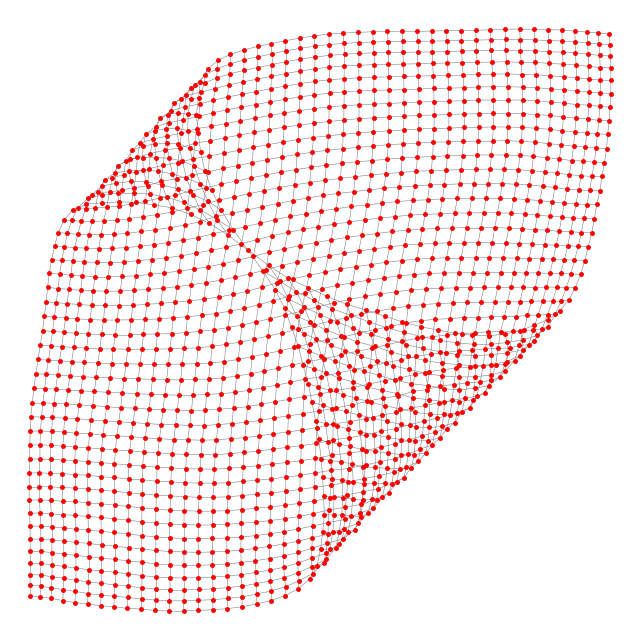}
        \end{minipage}
        \\\ 
      & $D_{3\_0}$ & $D_{3\_1}$ & $D_{3\_2}$ & $D_{3\_3}$ & $D_{3\_4}$ \\\
        \tabincell{l}{$G_{3}$\\high-density \\large graph\\$|V|$=2,851\\$|E|$=15,093\\$D$=5.29}
        &
          \begin{minipage}{0.15\textwidth}
            \includegraphics[width=\textwidth]{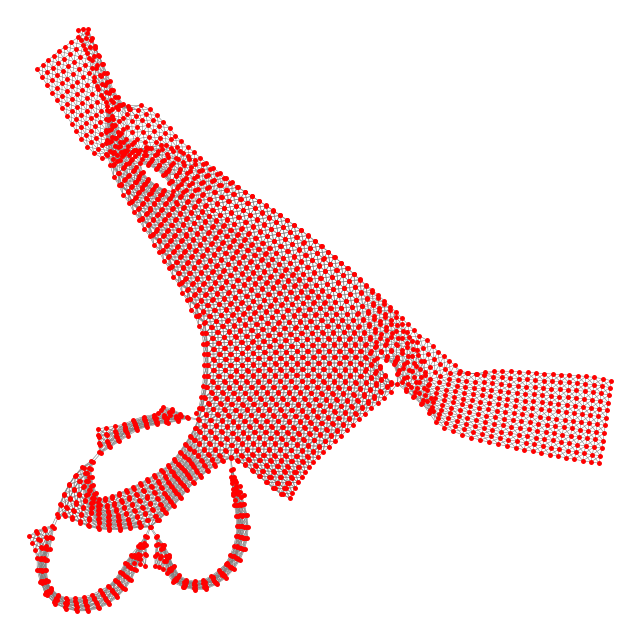}
        \end{minipage}
        &
        \begin{minipage}{0.15\textwidth}
            \includegraphics[width=\textwidth]{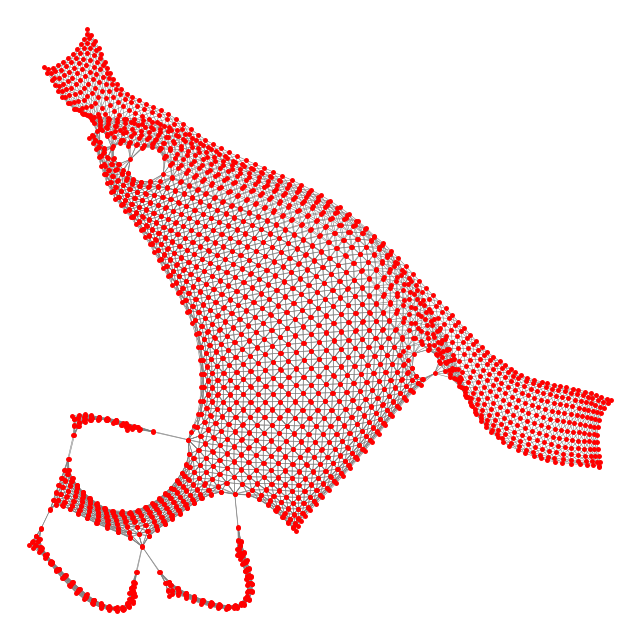}
        \end{minipage}
        &
        \begin{minipage}{0.15\textwidth}
            \includegraphics[width=\textwidth]{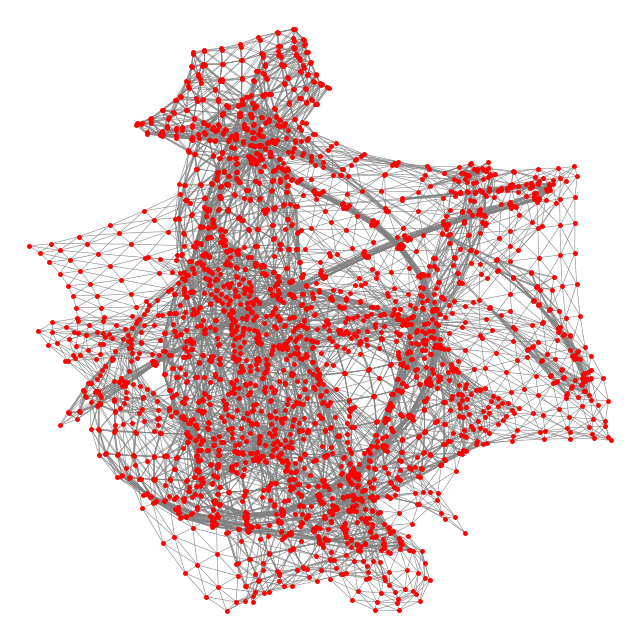}
        \end{minipage}
        &
        \begin{minipage}{0.15\textwidth}
            \includegraphics[width=\textwidth]{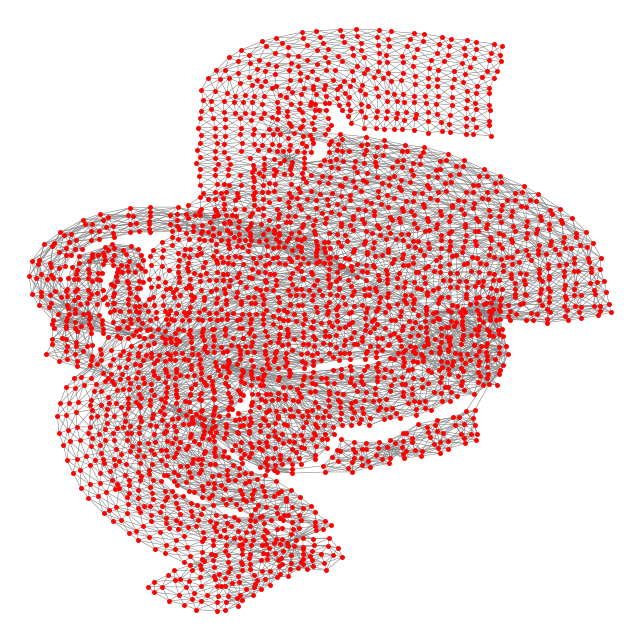}
        \end{minipage}
        &
        \begin{minipage}{0.15\textwidth}
            \includegraphics[width=\textwidth]{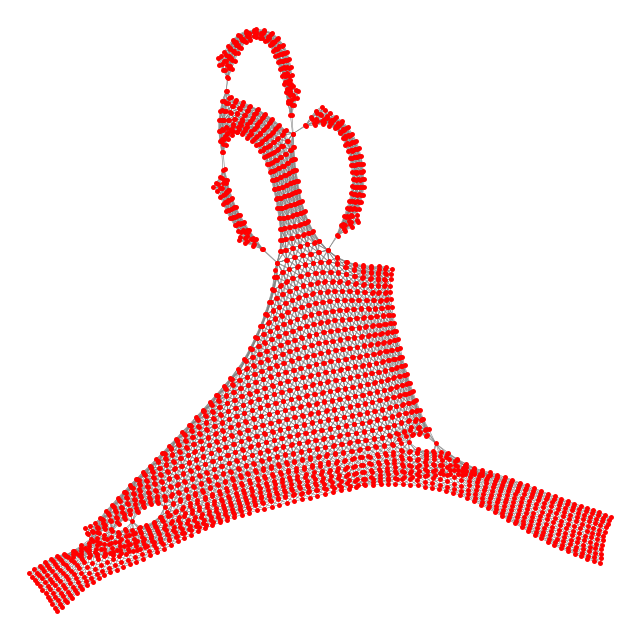}
        \end{minipage}
        \vspace{1mm}
        \\\
      & $D_{15\_0}$ & $D_{15\_1}$ & $D_{15\_2}$ & $D_{15\_3}$ & $D_{15\_4}$ \\\
        \tabincell{l}{$G_{15}$\\high-density \\large graph\\$|V|$=1,785\\$|E|$=20,459\\$D$=11.46}
        &
          \begin{minipage}{0.15\textwidth}
            \includegraphics[width=\textwidth]{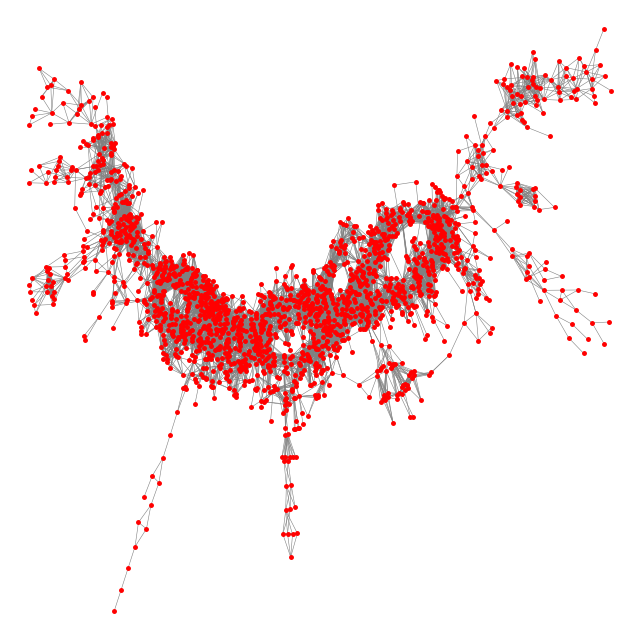}
        \end{minipage}
        &
        \begin{minipage}{0.15\textwidth}
            \includegraphics[width=\textwidth]{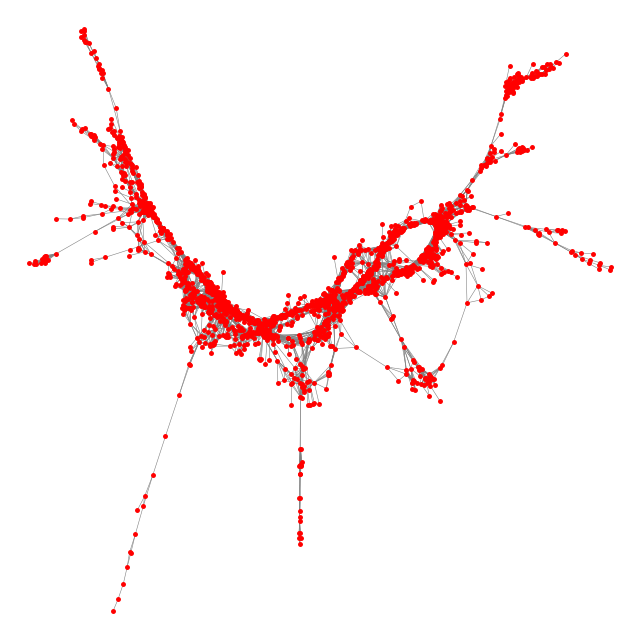}
        \end{minipage}
        &
        \begin{minipage}{0.15\textwidth}
            \includegraphics[width=\textwidth]{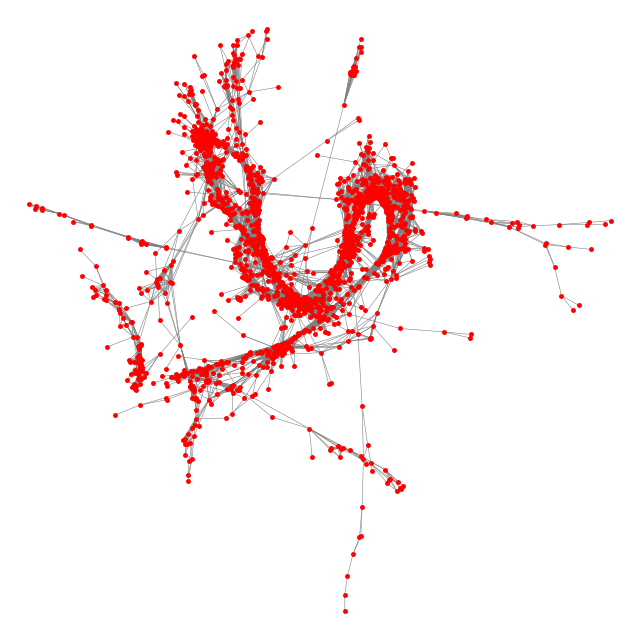}
        \end{minipage}
        &
        \begin{minipage}{0.15\textwidth}
            \includegraphics[width=\textwidth]{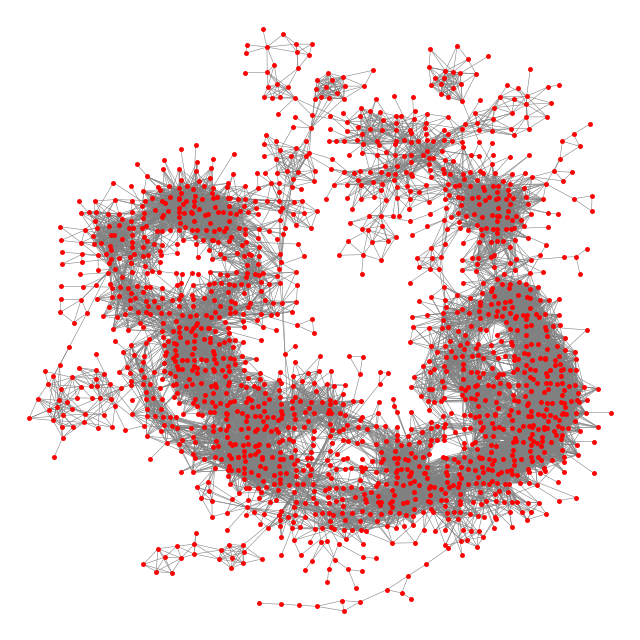}
        \end{minipage}
        &
        \begin{minipage}{0.15\textwidth}
            \includegraphics[width=\textwidth]{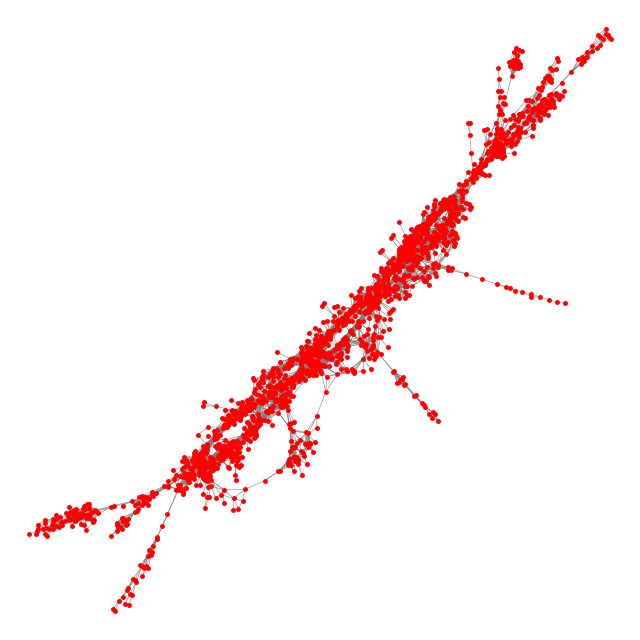}
        \end{minipage}
        \\\

\end{tabular} 
\caption{Examples of large graphs ($G_{6}$, $G_{10}$, $G_{3}$ and $G_{15}$) with their five different layouts.}
\label{pairs_mh}
\vspace{15mm}
\end{figure*}

\begin{figure*}[h]
\newcommand{\tabincell}[2]{\begin{tabular}{@{}#1@{}}#2\end{tabular}}
  \centering
    \begin{tabular}{l c c c c c}
      & $D_{0\_0}$ & $D_{0\_1}$ & $D_{0\_2}$ & $D_{0\_3}$ & $D_{0\_4}$
      \\\
        \tabincell{l}{$G_{0}$\\sparse\\$|V|$=363\\$|E|$=468\\$D$=1.29}
        &
          \begin{minipage}{0.15\textwidth}          
            \includegraphics[width=\textwidth]{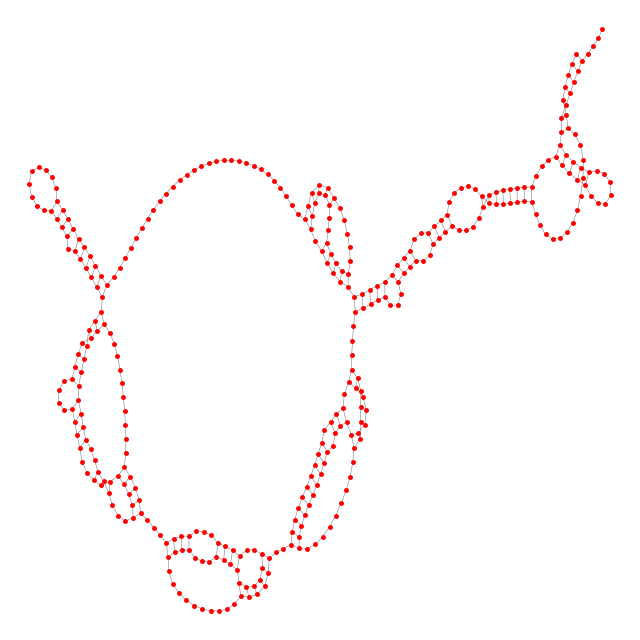}
        \end{minipage}
        &
        \begin{minipage}{0.15\textwidth}
            \includegraphics[width=\textwidth]{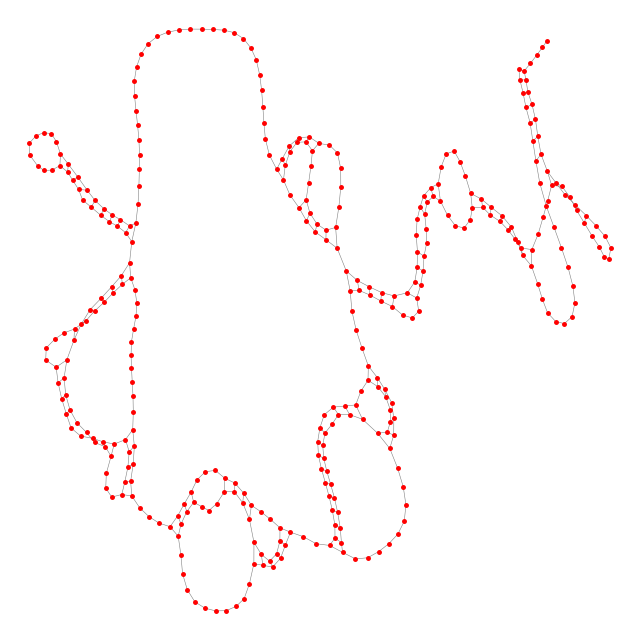}
        \end{minipage}
        &
        \begin{minipage}{0.15\textwidth}
            \includegraphics[width=\textwidth]{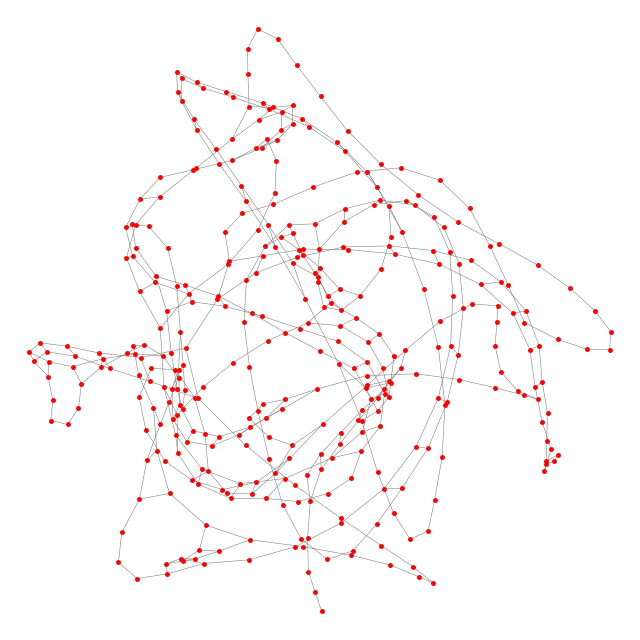}
        \end{minipage}
        &
        \begin{minipage}{0.15\textwidth}
            \includegraphics[width=\textwidth]{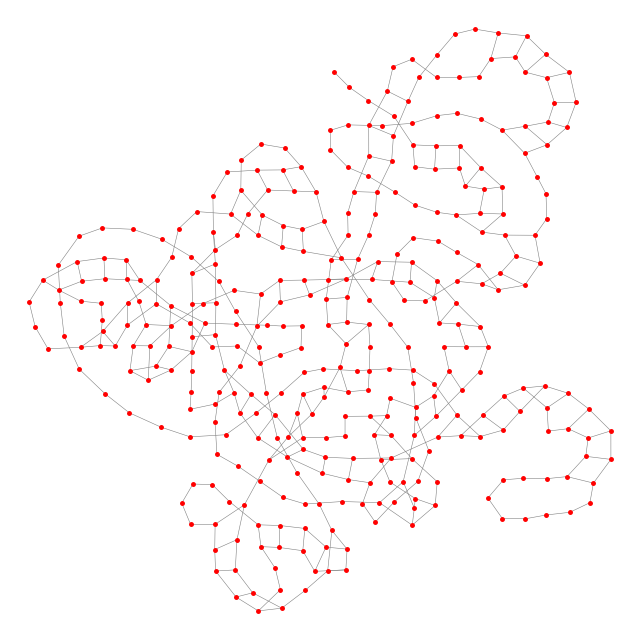}
        \end{minipage}
        &
        \begin{minipage}{0.15\textwidth}
            \includegraphics[width=\textwidth]{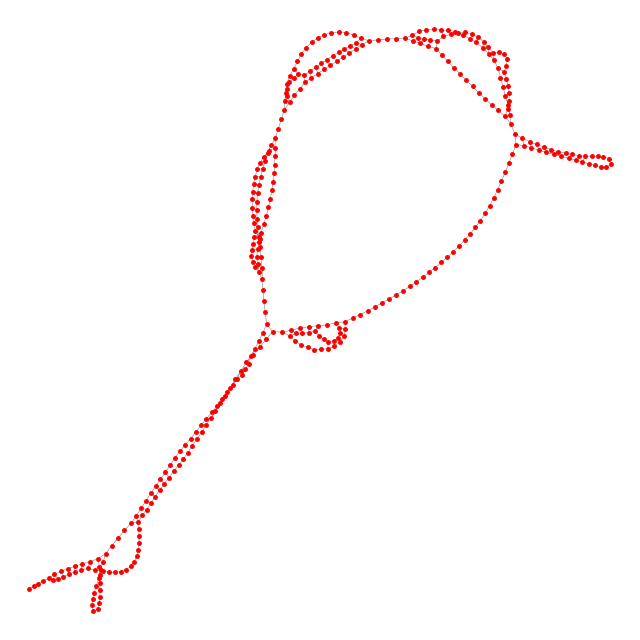}
        \end{minipage}
        \\\ 
      & $D_{188\_0}$ & $D_{188\_1}$ & $D_{188\_2}$ & $D_{188\_3}$ & $D_{188\_4}$ \\\
        \tabincell{l}{$G_{188}$\\sparse\\$|V|$=173\\$|E|$=181\\$D$=1.05}
        & 
          \begin{minipage}{0.15\textwidth}          
            \includegraphics[width=\textwidth]{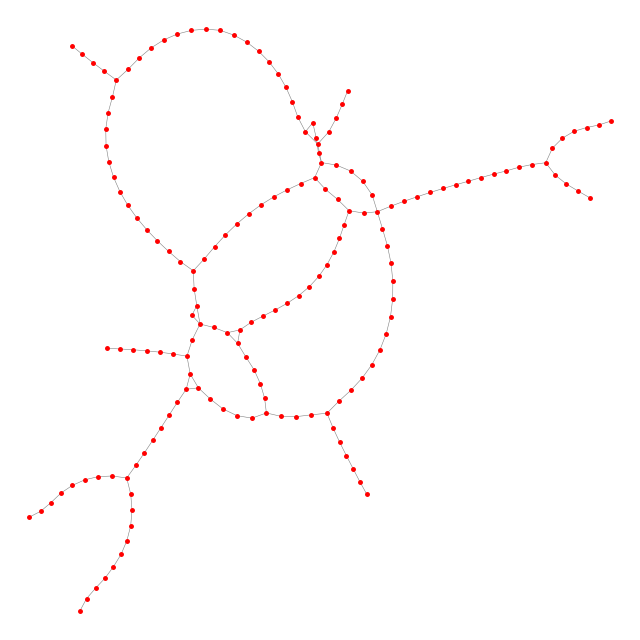}
        \end{minipage}
        &
        \begin{minipage}{0.15\textwidth}
            \includegraphics[width=\textwidth]{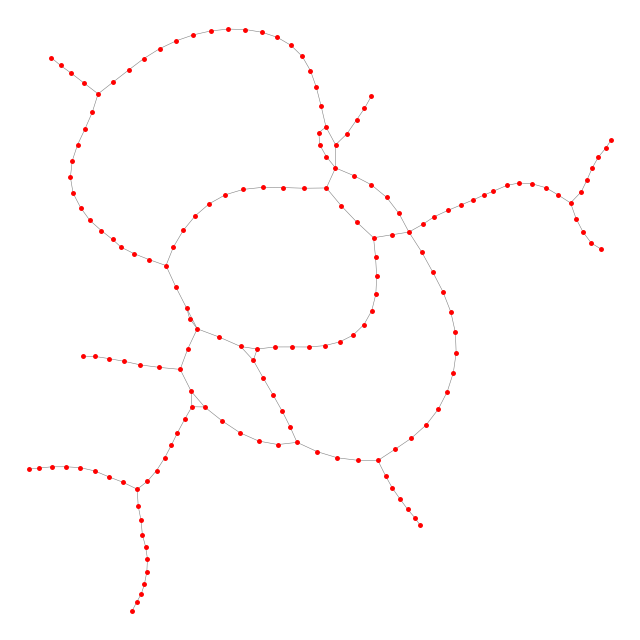}
        \end{minipage}
        &
        \begin{minipage}{0.15\textwidth}
            \includegraphics[width=\textwidth]{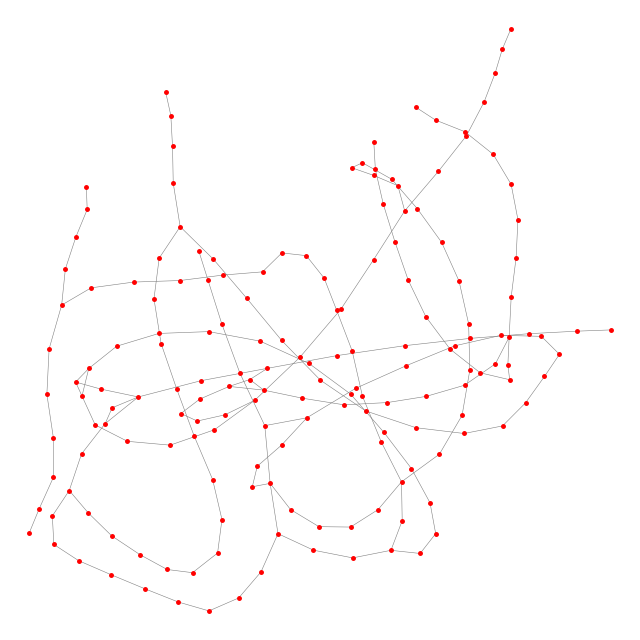}
        \end{minipage}
        &
        \begin{minipage}{0.15\textwidth}
            \includegraphics[width=\textwidth]{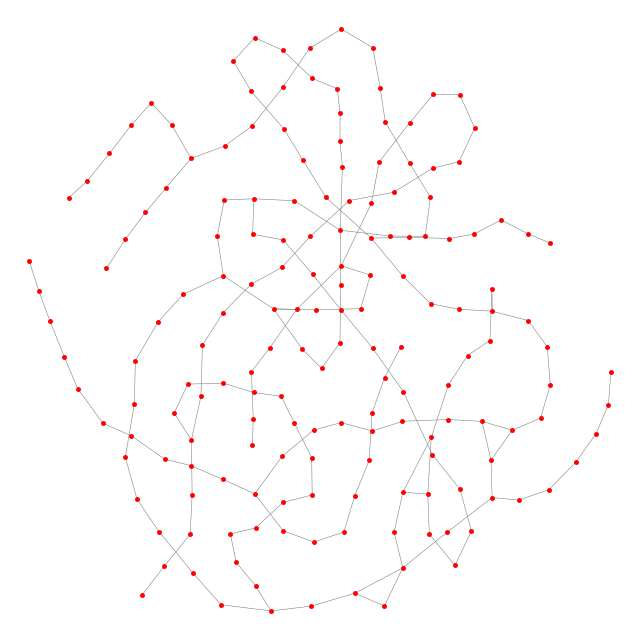}
        \end{minipage}
        &
        \begin{minipage}{0.15\textwidth}
            \includegraphics[width=\textwidth]{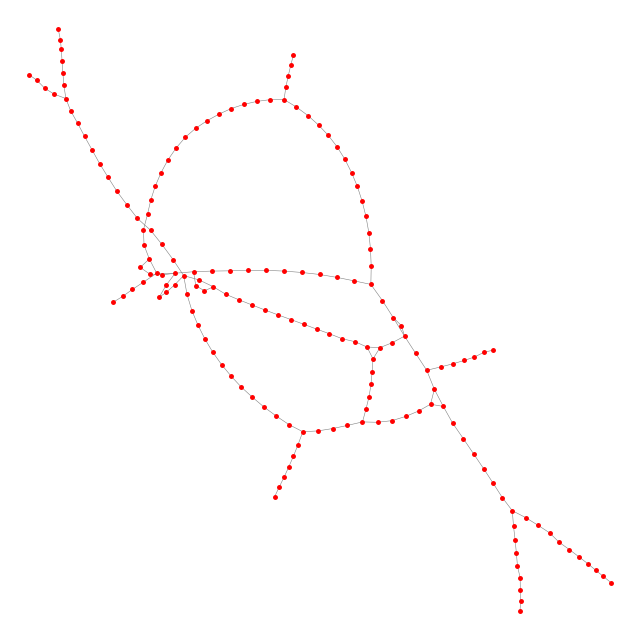}
        \end{minipage}
        \vspace{1mm}
        \\\
      & $D_{18\_0}$ & $D_{18\_1}$ & $D_{18\_2}$ & $D_{18\_3}$ & $D_{18\_4}$ \\\
        \tabincell{l}{$G_{18}$\\biconnected \\ graph\\$|V|$=70\\$|E|$=137\\$D$=1.96}
        &
          \begin{minipage}{0.15\textwidth}          
            \includegraphics[width=\textwidth]{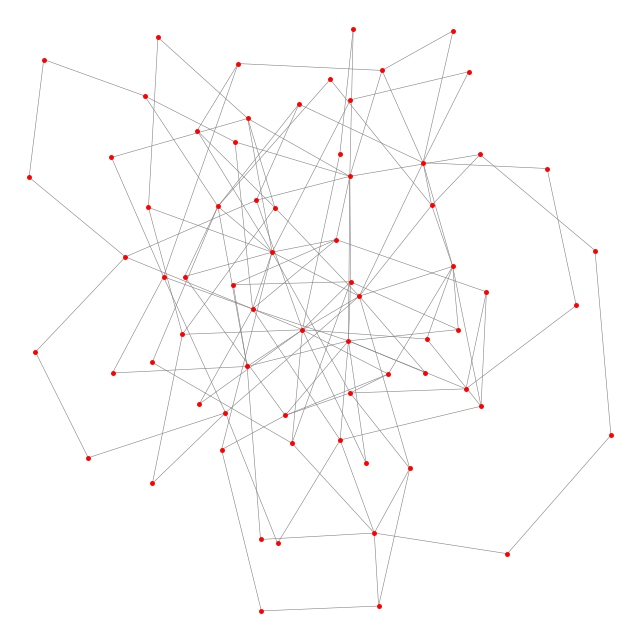}
        \end{minipage}
        &
        \begin{minipage}{0.15\textwidth}
            \includegraphics[width=\textwidth]{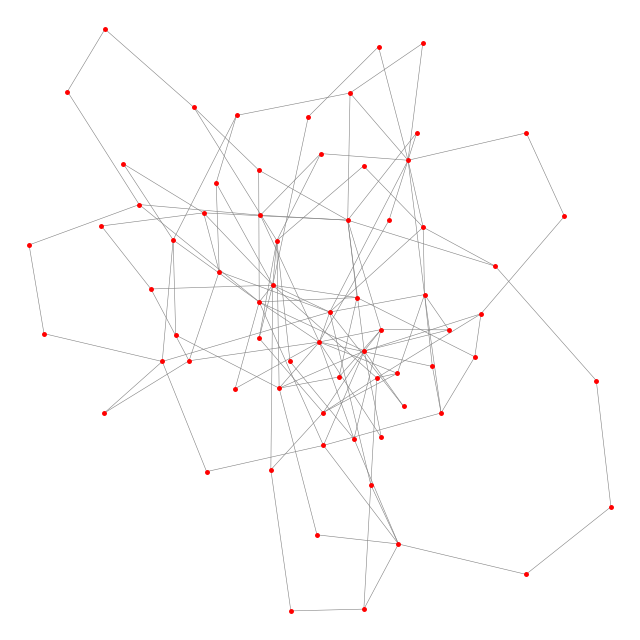}
        \end{minipage}
        &
        \begin{minipage}{0.15\textwidth}
            \includegraphics[width=\textwidth]{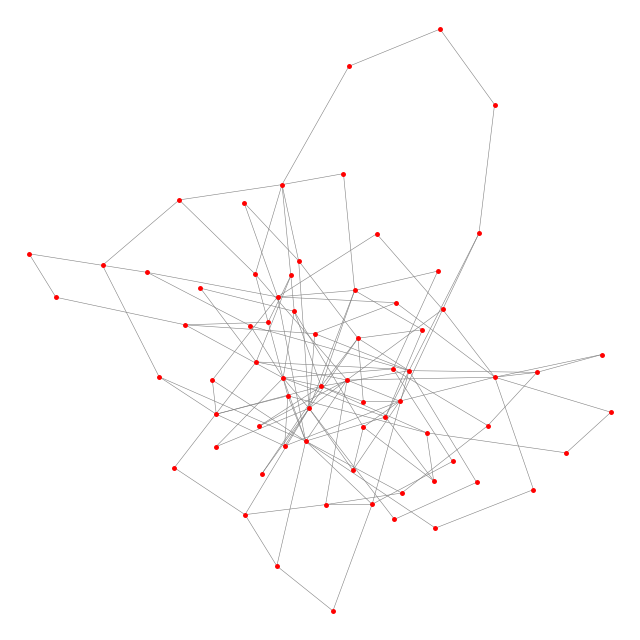}
        \end{minipage}
        &
        \begin{minipage}{0.15\textwidth}
            \includegraphics[width=\textwidth]{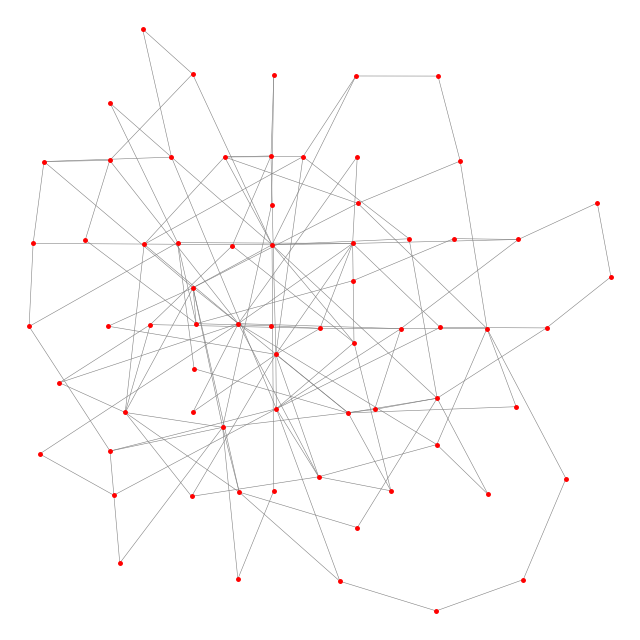}
        \end{minipage}
        &
        \begin{minipage}{0.15\textwidth}
            \includegraphics[width=\textwidth]{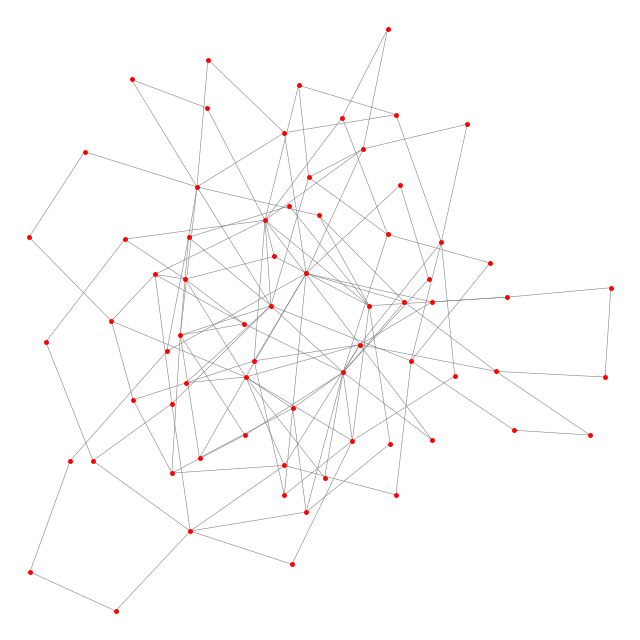}
        \end{minipage}
        \vspace{1mm}
        \\\
      & $D_{65\_0}$ & $D_{65\_1}$ & $D_{65\_2}$ & $D_{65\_3}$ & $D_{65\_4}$ \\\
        \tabincell{l}{$G_{65}$\\biconnected \\ graph\\$|V|$=230\\$|E|$=455\\$D$=1.98}
        &
          \begin{minipage}{0.15\textwidth}          
            \includegraphics[width=\textwidth]{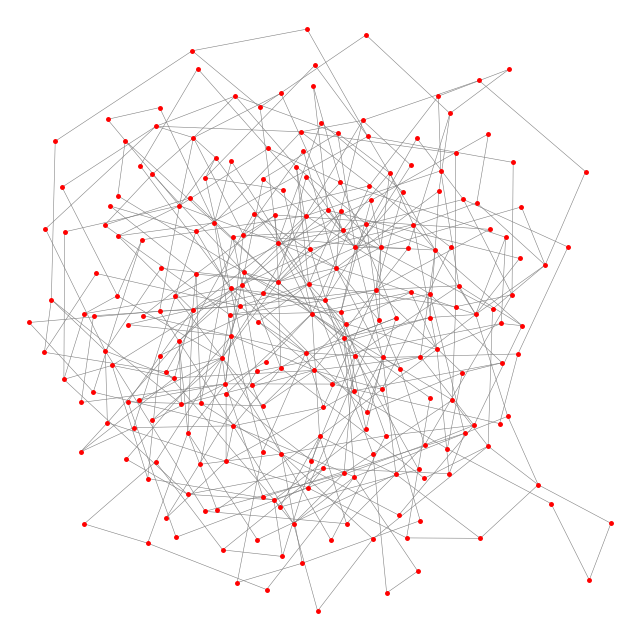}
        \end{minipage}
        &
        \begin{minipage}{0.15\textwidth}
            \includegraphics[width=\textwidth]{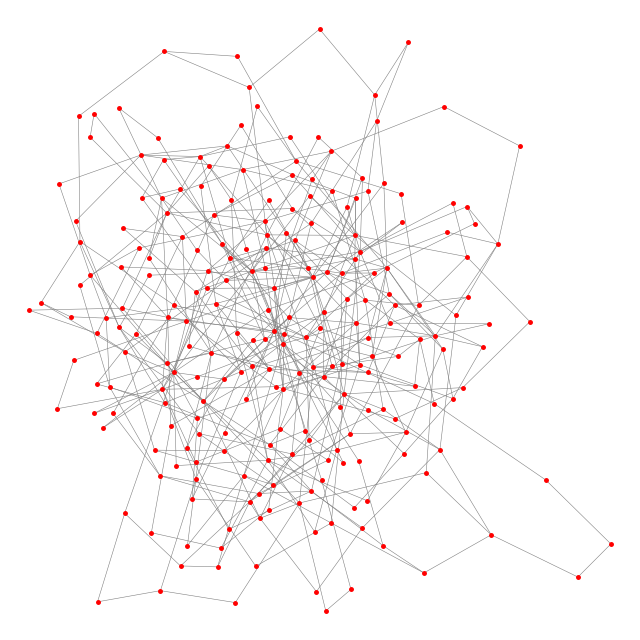}
        \end{minipage}
        &
        \begin{minipage}{0.15\textwidth}
            \includegraphics[width=\textwidth]{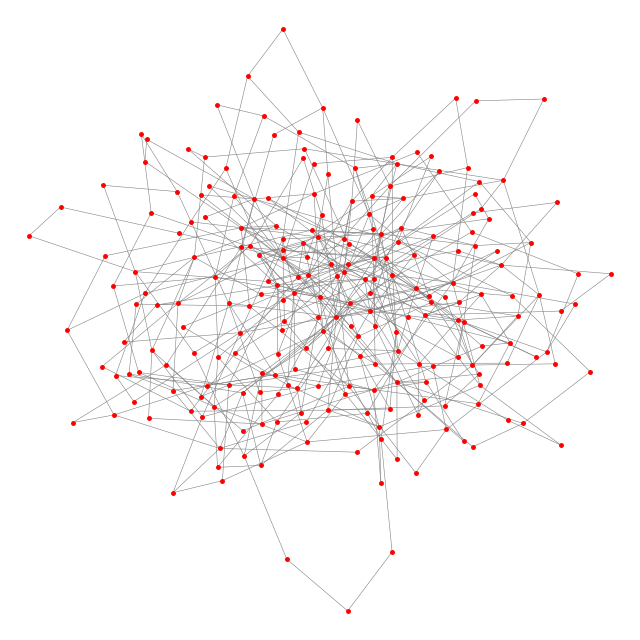}
        \end{minipage}
        &
        \begin{minipage}{0.15\textwidth}
            \includegraphics[width=\textwidth]{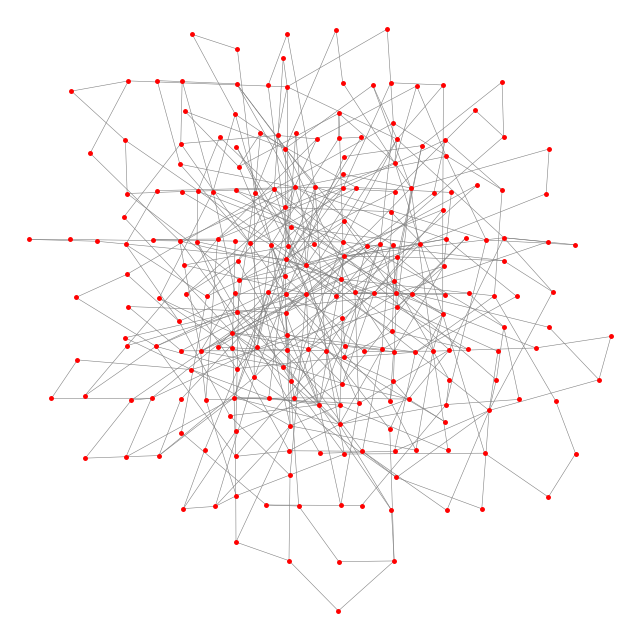}
        \end{minipage}
        &
        \begin{minipage}{0.15\textwidth}
            \includegraphics[width=\textwidth]{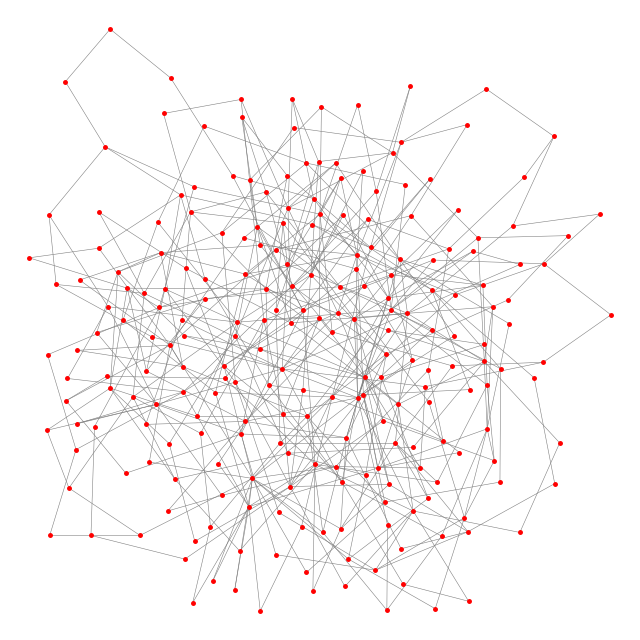}
        \end{minipage}
        \\\

\end{tabular}
\caption{Examples of small graphs ($G_{0}$, $G_{188}$, $G_{18}$ and $G_{65}$) with their five different layouts.} 
\label{pairs_cb}
\end{figure*}

\autoref{pairs_mh} and \autoref{pairs_cb} show the examples of four types of graphs and their five layouts used in the human preference experiments. 
\autoref{pairs_mh} shows the examples of large graphs $G_{6}$, $G_{10}$, $G_{3}$ and $G_{15}$.
\autoref{pairs_cb} shows the examples of small graphs $G_{0}$, $G_{188}$, $G_{18}$ and $G_{65}$.

In the previous experiments~\cite{eades2015shape}, the task for participants was to randomly choose from a pair of different layouts of the same graph, and then rate their preferred layout from 0 to 5. In our experiment, we convert the task performance from the preference score into human preference labels (e.g., 0 or 1) by employing the weighted majority voting algorithm.

Participants in the previous two experiments~\cite{eades2015shape} include computer science and graph drawing experts at the Graph Drawing conference in 2014, 
and students of the course information visualization at the University of Sydney.

\clearpage
\section{Detailed Discussion}\label{sec:further_exp}
\subsection{Comparison of trained models}\label{sec:Comparison_on_trained_models}

The model with fine-tuning by human preference pairs (i.e., M+HP) performs significantly better than the model without fine-tuning (i.e., M), and the model only exploiting the data with human preference labels (i.e., HP). 

The comparison results of M+HP vs M and M+HP vs HP show that the transfer learning technique is very useful and that the quality-metrics-based pairs contain knowledge that can help to predict human preference, consistent with the known correlation results~\cite{chimani2014people,eades2015shape}.

The success of transfer learning indicates that both the quality-metrics-based pairs and the human preference pairs are important in studying and predicting human preference for graph layouts.

The model trained by human preference pairs (i.e., HP) performs better than trained by quality-metrics-based pairs (i.e., M). 
This shows that the quality-metrics-based pairs are indeed different from the human preference pairs. 
Therefore, research on human preference for graph layouts should be investigated further.

\subsection{Large graphs}\label{sec:Large_graphs}

\begin{figure}[!h]
\newcommand{\tabincell}[2]{\begin{tabular}{@{}#1@{}}#2\end{tabular}}
  \centering
    \begin{tabular}{c c l l}
      $D_{5\_4}$ & $D_{5\_3}$ & & 
      \\\
      \begin{minipage}{0.16\textwidth}          
            \includegraphics[width=\textwidth]{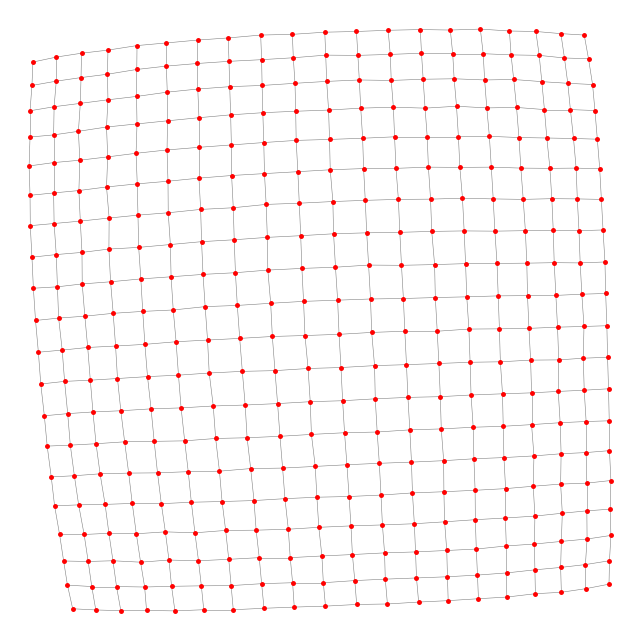}
        \end{minipage}
        &
        \begin{minipage}{0.16\textwidth}
            \includegraphics[width=\textwidth]{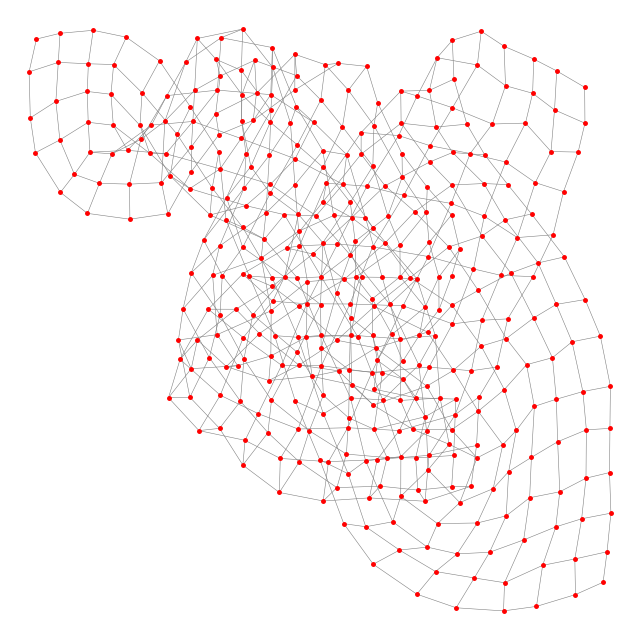}
        \end{minipage}
        &
        \small
        \tabincell{l}{$M$\\$HP$\\$M+HP$}
        \hspace{-4mm}
        &
        \small
        \tabincell{l}{$\surd$\\$\surd$\\$\surd$}
        \\\
         $D_{3\_0}$ & $D_{3\_3}$ & & 
      \\\      
      \begin{minipage}{0.16\textwidth}          
            \includegraphics[width=\textwidth]{G/G_3_0.png}
        \end{minipage}
        &
        \begin{minipage}{0.16\textwidth}
            \includegraphics[width=\textwidth]{G/G_3_3.png}
        \end{minipage}
        &
        \small
        \tabincell{l}{$M$\\$HP$\\$M+HP$}
        \hspace{-4mm}
        &
        \small
        \tabincell{l}{$\surd$\\$\surd$\\$\surd$}
        \\\
        \vspace{-6mm}
\end{tabular}
\caption{Examples of the test layout pairs for mesh $G_{5}$ and large scale-free graph $G_{3}$, where our three trained models succeed ($\surd$) to predict the ground truth human preference labels. In each row, the layout on the left is more preferred by human than layout on the right.} 
\label{pairs:g1}
\end{figure}

For mesh and large scale-free graphs, the test accuracy is very high, i.e., ($86.55 \pm 3.2$)\% and ($98 \pm 4.47$)\%, respectively. 
For instance, \autoref{pairs:g1} shows two examples that our three trained models (i.e., M, HP and M+HP) succeed to predict the ground truth human preference.

To compare the prediction results based on layouts, we further examine the five layouts of large graphs in detail. 
From the five layouts of $G_{6}$, $G_{10}$, $G_{3}$ and $G_{15}$ in~\autoref{pairs_mh}, we can see that layouts $D_{i\_0}$, $D_{i\_1}$ and $D_{i\_4}$ are all visually much better than layouts $D_{i\_2}$ and $D_{i\_3}$.
Therefore, it was easy for human to make a preference from a pair between $D_{i\_0}$ (resp. $D_{i\_1}$ and $D_{i\_4}$) and $D_{i\_2}$ (resp. $D_{i\_3}$) (see~\autoref{pairs:g1}).

\begin{figure}[!h]
\newcommand{\tabincell}[2]{\begin{tabular}{@{}#1@{}}#2\end{tabular}}
  \centering
    \begin{tabular}{c c l l}
    $D_{7\_0}$ & $D_{7\_1}$ & & 
      \\\
      \begin{minipage}{0.16\textwidth}         
            \includegraphics[width=\textwidth]{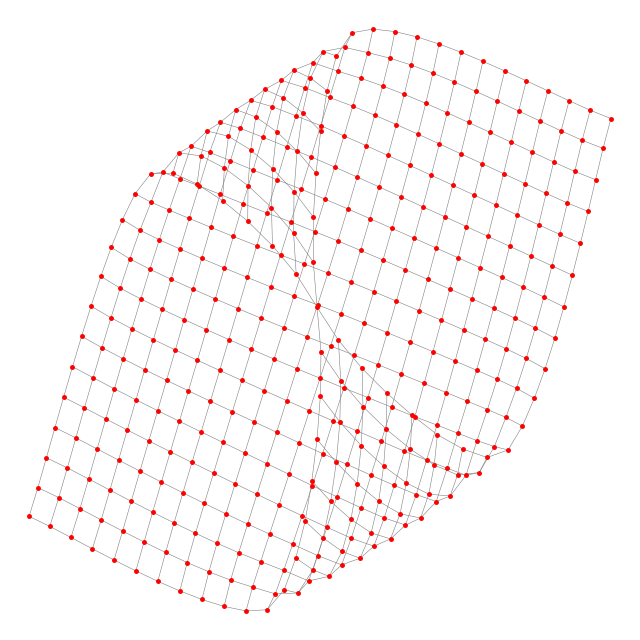}
        \end{minipage}
        &
        \begin{minipage}{0.16\textwidth}
            \includegraphics[width=\textwidth]{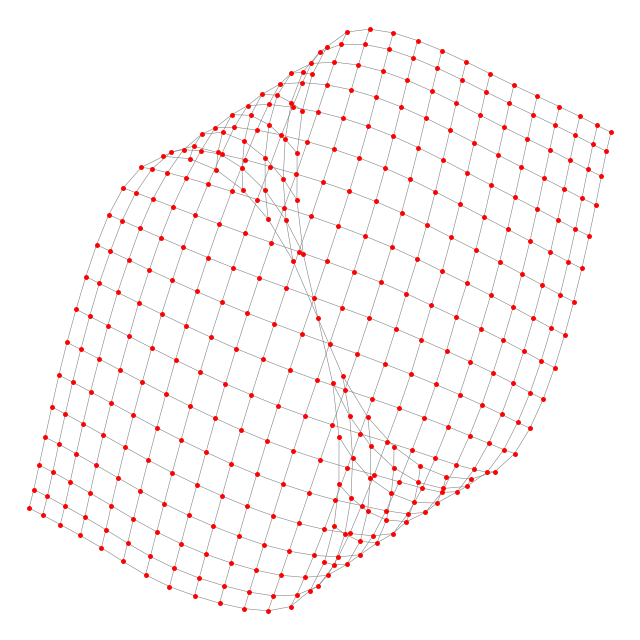}
        \end{minipage}
        &
        \small
        \tabincell{l}{$M$\\$HP$\\$M+HP$}
        \hspace{-4mm}
        &
        \small
        \tabincell{l}{$\times$\\$\surd$\\$\surd$}
        \\\
        $D_{3\_3}$ & $D_{3\_2}$ & & 
      \\\      
      \begin{minipage}{0.16\textwidth}         
            \includegraphics[width=\textwidth]{G/G_3_3.png}
        \end{minipage}
        &
        \begin{minipage}{0.16\textwidth}
            \includegraphics[width=\textwidth]{G/G_3_2.png}
        \end{minipage}
        &
        \small
        \tabincell{l}{$M$\\$HP$\\$M+HP$}
        \hspace{-4mm}
        &
        \small
        \tabincell{l}{$\times$\\$\surd$\\$\surd$}
        \\\
        \vspace{-6mm}
\end{tabular}
\caption{Examples of  the test layout pairs for mesh $G_{7}$ and large scale-free graph $G_{3}$, where our three trained models succeed ($\surd$) or fail ($\times$) to predict the ground truth human preference labels. In each row, the layout on the left is more preferred by human than layout on the right.} 
\label{pairs:e1}
\vspace{-3mm}
\end{figure}

On the other hand, there are difficult cases for human to decide which layout they prefer, i.e., layouts $D\_i\_0$ and $D\_i\_1$ are similar and slightly better than $D\_i\_4$.
In this case, a layout was chosen with conflicts
(e.g., for the same layout pair [$D\_7\_0$, $D\_7\_1$], the preference among three participants was different,
see~\autoref{pairs:e1} and~\autoref{e1}).

Similarly, layouts $D\_i\_2$ and $D\_i\_3$ are of similar quality, both worse than the other three layouts, difficult for human to make a selection.
In this case, a layout was chosen with a very low preference score (see  [$D\_3\_2$, $D\_3\_3$] in~\autoref{pairs:e1} and~\autoref{e1}).

\begin{table}[!h]
\centering
    \begin{tabular}{|l|c|c|c|}
    \hline
$D_{i\_j}$&$D_{i\_k}$&$P$&$S$\\\hline
$D_{7\_0}$&$D_{7\_1}$&$D_{7\_1}$&$1$\\\hline
$D_{7\_0}$&$D_{7\_1}$&$D_{7\_0}$&$4$\\\hline
$D_{7\_1}$&$D_{7\_0}$&$D_{7\_0}$&$3$\\\hline
$D_{3\_3}$&$D_{3\_2}$&$D_{3\_3}$&$1$\\\hline
    \end{tabular}
    \vspace{2mm}
 \caption{Examples of the preference and preference score of participants for choosing a preferred layout from layout pairs [$D_{7\_0}$, $D_{7\_1}$] and [$D_{3\_2}$, $D_{3\_3}$].}
 \label{e1}
\vspace{-6mm}
\end{table}

\subsection{Small graphs}\label{sec:Small_graphs}
\begin{figure}[!h]
\newcommand{\tabincell}[2]{\begin{tabular}{@{}#1@{}}#2\end{tabular}}
  \centering
    \begin{tabular}{c c l l}
    $D_{185\_0}$ & $D_{185\_3}$ & & 
      \\\
      \begin{minipage}{0.16\textwidth}          
            \includegraphics[width=\textwidth]{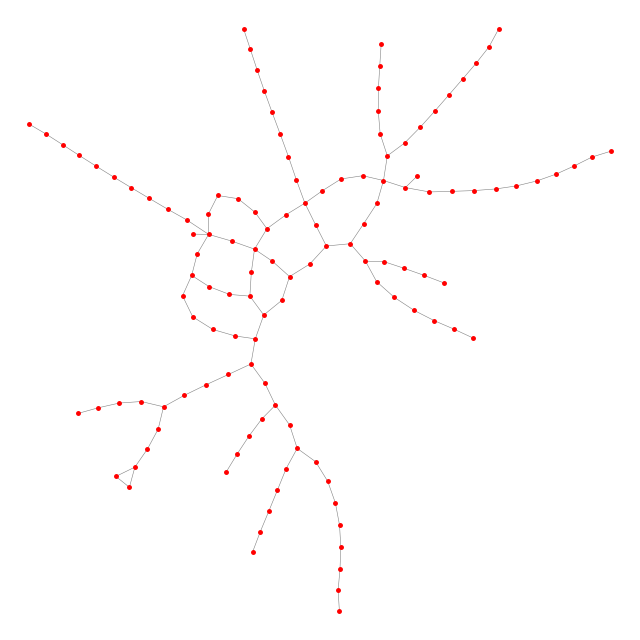}
        \end{minipage}
        &
        \begin{minipage}{0.16\textwidth}
            \includegraphics[width=\textwidth]{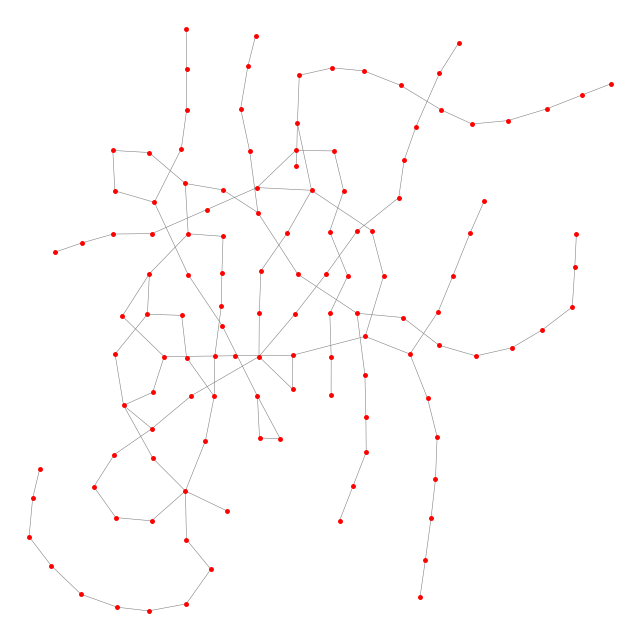}
        \end{minipage}
        &
        \small
        \tabincell{l}{$M$\\$HP$\\$M+HP$}
        \hspace{-4mm}
        &
        \small
        \tabincell{l}{$\surd$\\$\surd$\\$\surd$}
      \\\
      $D_{42\_0}$ & $D_{42\_2}$ & & 
      \\\
      \begin{minipage}{0.16\textwidth}          
            \includegraphics[width=\textwidth]{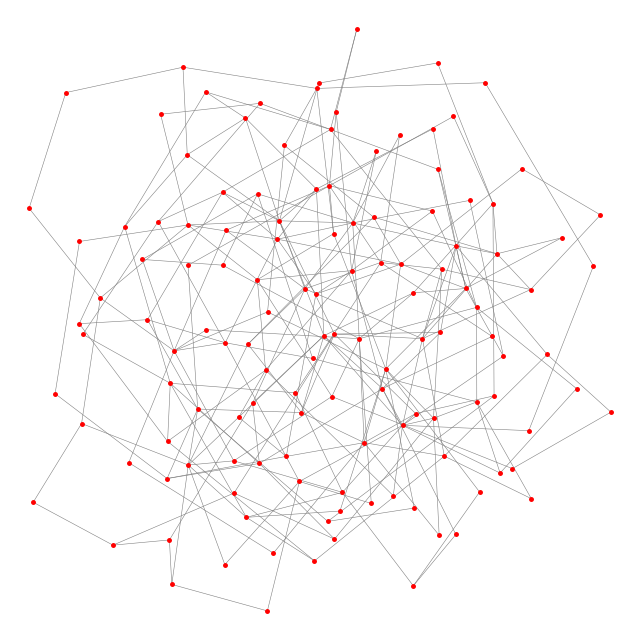}
        \end{minipage}
        &
        \begin{minipage}{0.16\textwidth}
            \includegraphics[width=\textwidth]{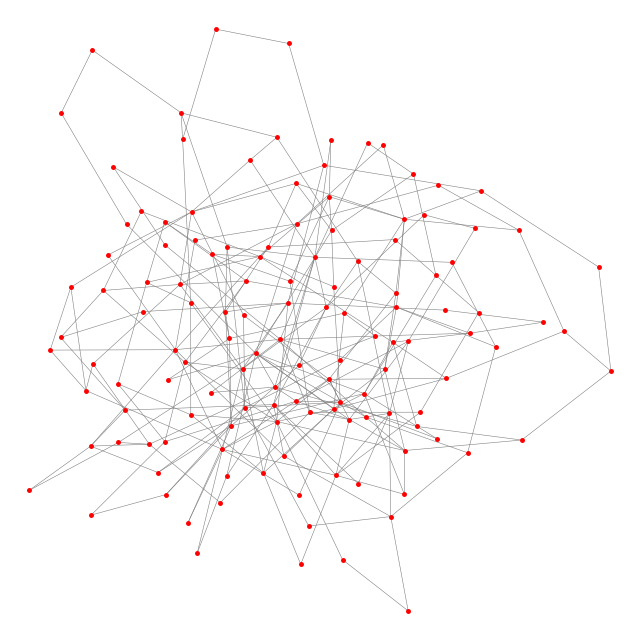}
        \end{minipage}
        &
        \small
        \tabincell{l}{$M$\\$HP$\\$M+HP$}
        \hspace{-4mm}
        &
        \small
        \tabincell{l}{$\surd$\\$\surd$\\$\surd$}
        \\\
        \vspace{-6mm}
\end{tabular}
\caption{Examples of the test layout pairs for sparse $G_{185}$ and biconnected graph $G_{42}$, where our three trained models succeed ($\surd$) to predict the ground truth human preference labels. In each row, the layout on the left is more preferred by human than layout on the right.} 
\label{pairs:g2}
\end{figure}

Overall, the test accuracy for small graphs is relatively low compared with large graphs. 
For instance, \autoref{pairs:g2} shows two examples of layout pairs of sparse graph $G_{185}$ and biconnected graph $G_{42}$.

\begin{figure}[!h]
\newcommand{\tabincell}[2]{\begin{tabular}{@{}#1@{}}#2\end{tabular}}
  \centering
    \begin{tabular}{c c l l}
  $D_{185\_0}$ & $D_{185\_4}$ & & 
      \\\
      \begin{minipage}{0.16\textwidth}          
            \includegraphics[width=\textwidth]{G/G_185_0.png}
        \end{minipage}
        &
        \begin{minipage}{0.16\textwidth}
            \includegraphics[width=\textwidth]{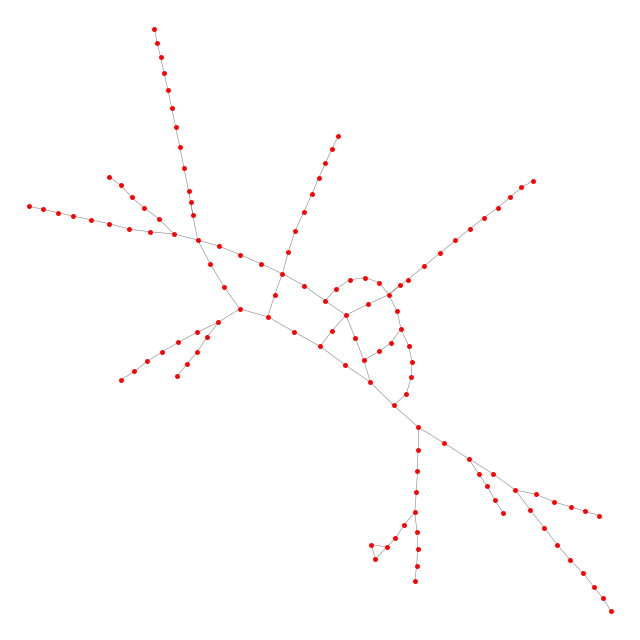}
        \end{minipage}
        &
        \small
        \tabincell{l}{$M$\\$HP$\\$M+HP$}
        \hspace{-4mm}
        &
        \small
        \tabincell{l}{$\times$\\$\surd$\\$\surd$}
        \\\
      $D_{66\_3}$ & $D_{66\_0}$ & & 
      \\\
      \begin{minipage}{0.16\textwidth}            
            \includegraphics[width=\textwidth]{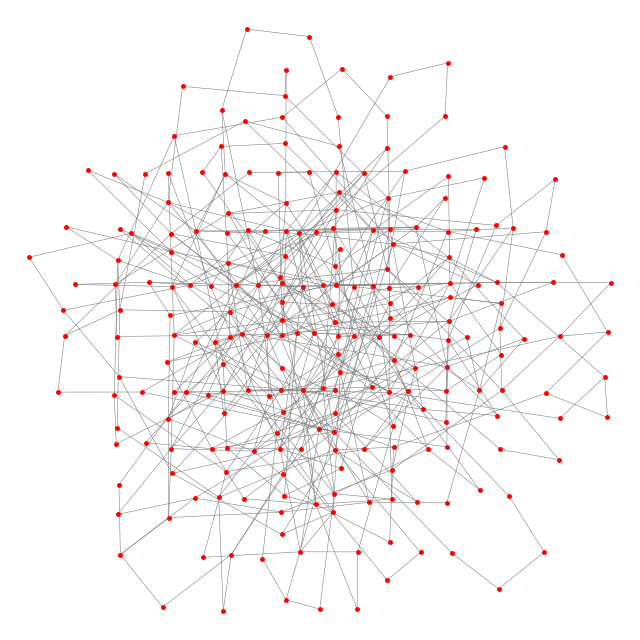}
        \end{minipage}
        &
        \begin{minipage}{0.16\textwidth}
            \includegraphics[width=\textwidth]{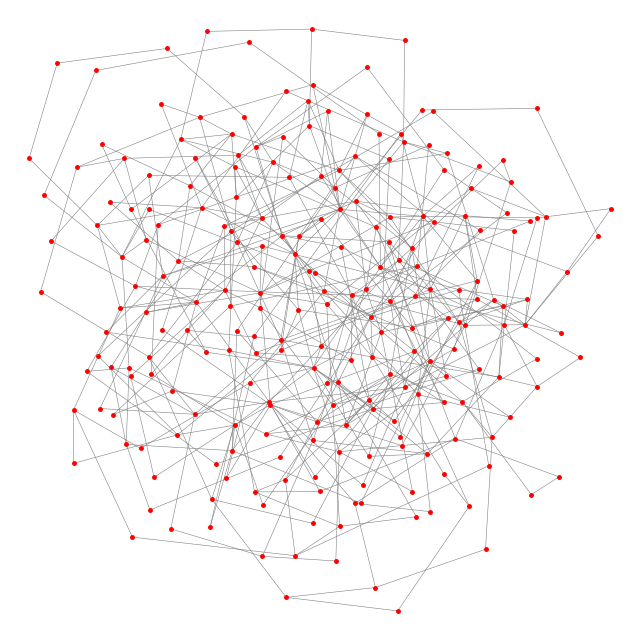}
        \end{minipage}
        &
        \small
        \tabincell{l}{$M$\\$HP$\\$M+HP$}
        \hspace{-4mm}
        &
        \small
        \tabincell{l}{$\times$\\$\times$\\$\surd$}
        \\\
        \vspace{-6mm}
\end{tabular}
\caption{Examples of the test layout pairs for sparse graph $G_{185}$ and biconnected graph $G_{66}$, where our three trained models succeed ($\surd$) or fail ($\times$) to predict the ground truth human preference labels. In each row, the layout on the left is more preferred by human than layout on the right.
} 
\label{pairs:e2}
\vspace{-3mm}
\end{figure}%
\begin{table}[!h]
\centering
    \begin{tabular}{|l|c|c|c|}
    \hline
$D_{i\_j}$&$D_{i\_k}$&$P$&$S$\\\hline
$D_{185\_0}$&$D_{185\_4}$&$D_{185\_0}$&$1$\\\hline
$D_{185\_0}$&$D_{185\_4}$&$D_{185\_0}$&$2$\\\hline
$D_{66\_0}$&$D_{66\_3}$&$D_{66\_3}$&$1$\\\hline
    \end{tabular}
    \vspace{2mm}
\caption{Examples of the preference and preference score of participants for choosing a preferred layout from layout pairs [$D_{185\_0}$, $D_{185\_4}$] and [$D_{66\_0}$, $D_{66\_3}$].}
 \label{e2}
\end{table}

The low test accuracy is because the layout pairs of small graphs are difficult for human to make a preference. 
This is because all the five graph layout algorithms produced visually similar layouts with similar quality for small graphs  (see sparse graph $G_{188}$, and biconnected graphs $G_{18}$ and $G_{65}$ in~\autoref{pairs_cb}). 
Therefore, human tend to randomly choose a layout as the preference with a low preference score, 
e.g., \autoref{pairs:e2} and \autoref{e2} show that the preference score by each participant was very low for both the sparse graph layout pair [$D_{185\_0}$, $D_{185\_4}$] and the biconnected graph layout pair [$D_{66\_0}$, $D_{66\_3}$].

\subsection{Limitation and future improvement}\label{sec:Limitation_and_future_improvement}

Overall our trained models perform quite well, nevertheless, there are some unexpected cases. 
We use three examples to address and analyze our limitations in detail.

\begin{figure}[!h]
\newcommand{\tabincell}[2]{\begin{tabular}{@{}#1@{}}#2\end{tabular}}
  \centering
    \begin{tabular}{c c l l}
    $D_{13\_4}$ & $D_{13\_0}$ & & 
      \\\      
      \begin{minipage}{0.16\textwidth}
            \includegraphics[width=\textwidth]{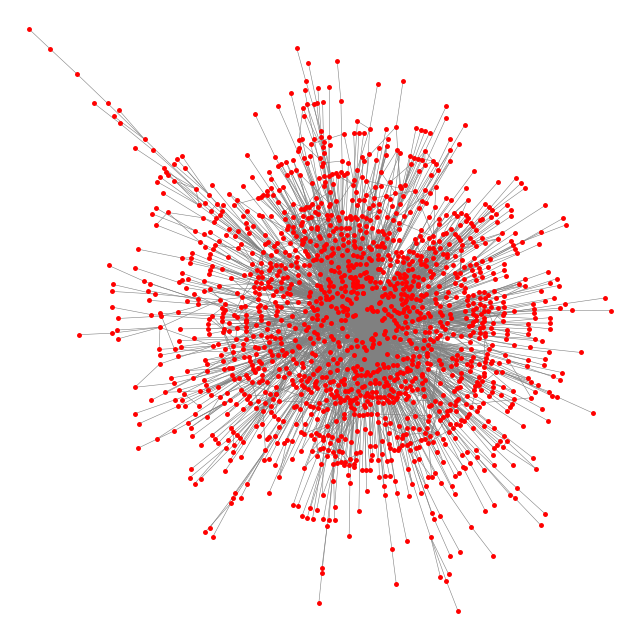}
        \end{minipage}
        &
        \begin{minipage}{0.16\textwidth}
            \includegraphics[width=\textwidth]{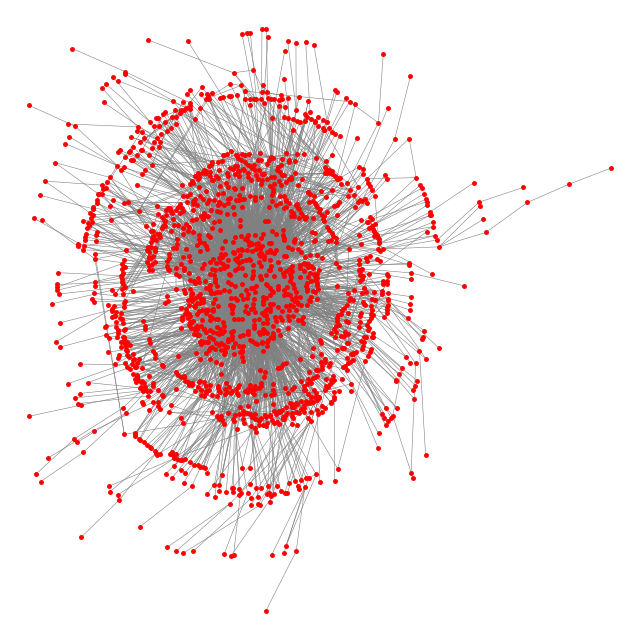}
        \end{minipage}
        &
        \small
        \tabincell{l}{$M$\\$HP$\\$M+HP$}
        \hspace{-4mm}
        &
        \small
        \tabincell{l}{$\times$\\$\times$\\$\times$}
        \\\
      $D_{113\_2}$ & $D_{113\_1}$ & & 
      \\\
      \begin{minipage}{0.16\textwidth}
            \includegraphics[width=\textwidth]{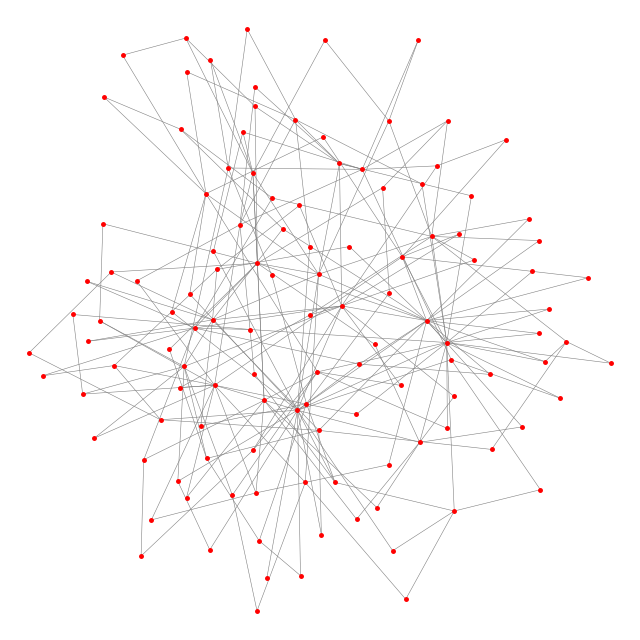}
        \end{minipage}
        &
        \begin{minipage}{0.16\textwidth}
            \includegraphics[width=\textwidth]{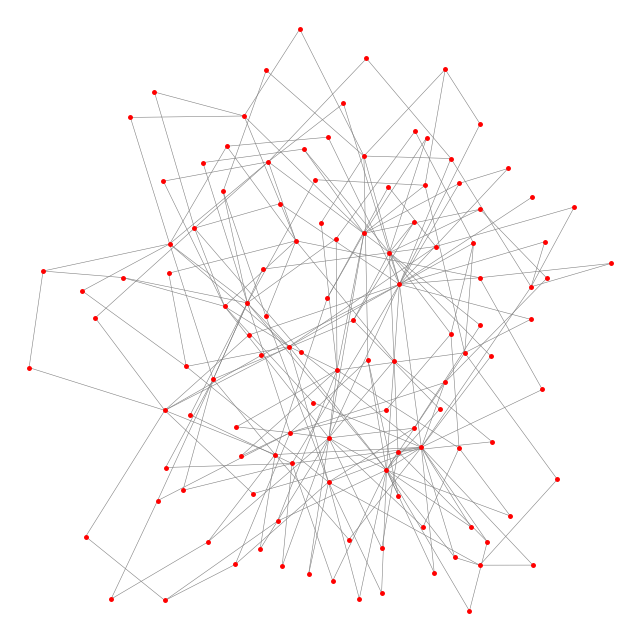}
        \end{minipage}
        &
        \small
        \tabincell{l}{$M$\\$HP$\\$M+HP$}
        \hspace{-4mm}
        &
        \small
        \tabincell{l}{$\times$\\$\times$\\$\times$}
        \\\
        \vspace{-6mm}
\end{tabular}
\caption{Examples of the test layout pairs for $G_{13}$, and $G_{113}$, where all our three trained models fail ($\times$) to predict the ground truth human preference labels. In each row, the layout on the left is more preferred by human than layout on the right.} 
\label{pairs:f}
\vspace{-6mm}
\end{figure}

\autoref{pairs:f} shows the instance where all our models fail to predict the ground truth human preference on the test data. 
\autoref{f} shows the corresponding preference score.

From~\autoref{pairs:f}, we can see that each layout pair has almost the same visual quality.
For example, two layouts of the pair [$D_{113\_1}$, $D_{113\_2}$],  are visually similar, and the preference score of preferred layout $D_{113\_2}$ is very low. 
Due to such a low preference score, our binary classification accuracy for biconnected graph is lower than large graphs.

Furthermore, human preference can be subjective depending on personal preference, resulting in different preference scores, e.g., preference scores of 1 and 4 for the same pair [$D_{13\_0}$, $D_{13\_4}$].

\begin{figure}[!h]
\newcommand{\tabincell}[2]{\begin{tabular}{@{}#1@{}}#2\end{tabular}}
  \centering
    \begin{tabular}{c c l l}
        $D_{187\_4}$ & $D_{187\_2}$ & & 
      \\\
      \begin{minipage}{0.16\textwidth}
            \includegraphics[width=\textwidth]{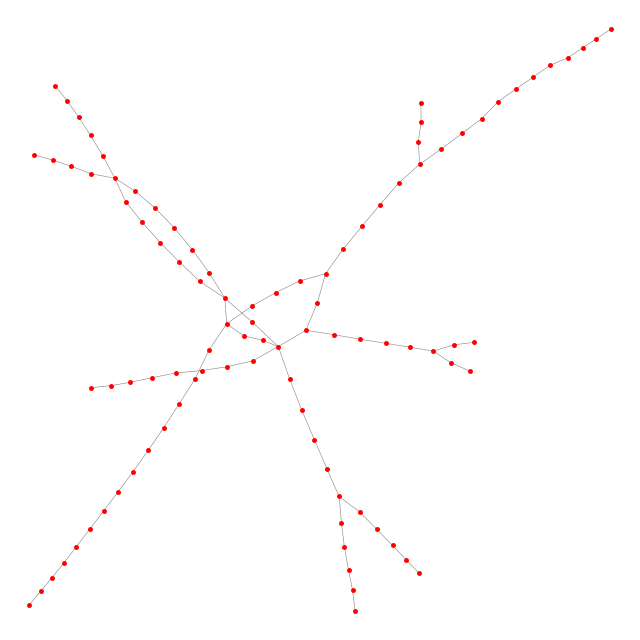}
        \end{minipage}
        &
        \begin{minipage}{0.16\textwidth}
            \includegraphics[width=\textwidth]{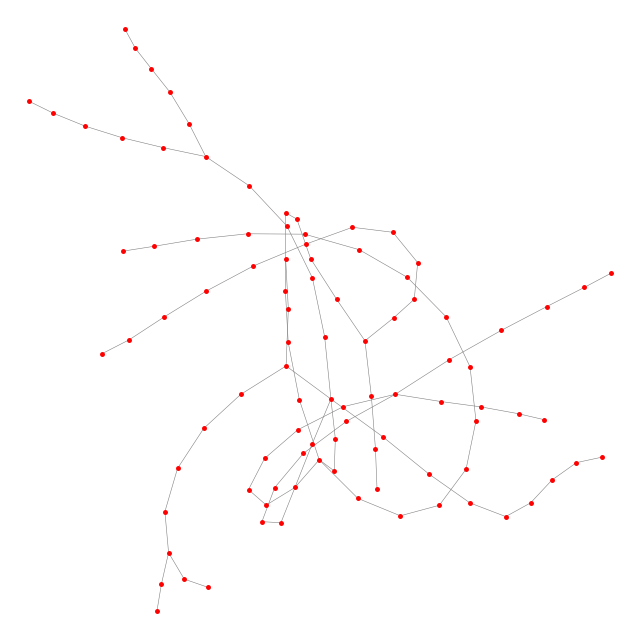}
        \end{minipage}
        &
        \small
        \tabincell{l}{$M$\\$HP$\\$M+HP$}
        \hspace{-4mm}
        &
        \small
        \tabincell{l}{$\surd$\\$\times$\\$\times$}
        \\\
        \vspace{-6mm}
\end{tabular}
\caption{Example of the test layout pair [$D_{187\_2}$, $D_{187\_4}$], where model M succeeds ($\surd$) (resp. models HP and M+HP fail ($\times$)) to predict the ground truth human preference label. The layout $D_{187\_4}$ is more preferred by human than layout $D_{187\_2}$.}
\label{pairs:f2}
\vspace{-8mm}
\end{figure}

On the other hand, \autoref{pairs:f2} and~\autoref{f2} show an example of an unexpected case, where only model M succeeds to predict the ground truth human preference. 
The preference score for the pair [$D_{187\_2}$, $D_{187\_4}$] is 4. 

\begin{table}[!h]
\centering
    \begin{tabular}{|l|c|c|c|}
    \hline
$D_{i\_j}$&$D_{i\_k}$&$P$&$S$\\\hline
$D_{13\_4}$&$D_{13\_0}$&$D_{13\_4}$&$1$\\\hline
$D_{13\_0}$&$D_{13\_4}$&$D_{13\_4}$&$4$\\\hline
$D_{113\_1}$&$D_{113\_2}$&$D_{113\_2}$&$1$\\\hline
    \end{tabular}
    \vspace{2mm}
\caption{Examples of the preference and preference score of participants for choosing a preferred layout from layout pairs [$D_{13\_0}$, $D_{13\_4}$] and [$D_{113\_1}$, $D_{113\_2}$].}
 \label{f}
\end{table}

\begin{table}[!h]
\centering
    \begin{tabular}{|l|c|c|c|}
    \hline
$D_{i\_j}$&$D_{i\_k}$&$P$&$S$\\\hline
$D_{187\_4}$&$D_{187\_2}$&$D_{187\_4}$&$4$\\\hline
    \end{tabular}
    \vspace{2mm}
 \caption{Example of the preference and preference score of participants for choosing a preferred layout from layout pairs [$D_{187\_2}$, $D_{187\_4}$] with the preference score 4.}
 \label{f2}
\vspace{-5mm}
\end{table}

\end{appendices}

\end{document}